\documentclass[journal]{IEEEtran}

\ifCLASSINFOpdf
\else
\fi

\usepackage{amsmath}
\usepackage{amssymb}
\usepackage{graphicx}
\usepackage{cite}

\begin{document}

\title{Wideband Near- and Far-Field Double-Bounce Physical Optics via Co-Area Range Profile Extraction}

\author{Guy~Binderman,~\IEEEmembership{Student Member,~IEEE,}
        and~Amir~Boag,~\IEEEmembership{Fellow,~IEEE}%
\thanks{G. Binderman and A. Boag are with the Department of Physical Electronics,
School of Electrical Engineering, Tel Aviv University, Tel Aviv, Israel.}}

\markboth{Preprint}%
{Binderman \MakeLowercase{\textit{et al.}}: Wideband Double-Bounce Physical Optics via Co-Area Range Profile Extraction}

\maketitle

\begin{abstract}
Wideband evaluation of the Radar Cross Section (RCS) and the underlying range profile of a target, including its secondary double-bounce fields, remains a computational bottleneck even for modern high-frequency solvers. This paper presents a framework for evaluating the wideband range profile under the Physical Optics (PO) approximation without repeated frequency-domain iterations. Using phase level-set integration via the Co-Area formula, we project the single- and double-bounce scattered fields into frequency-independent Geometric Range Profiles (GRPs), valid in both the far-field and near-field regions. The approach builds on the Radon Transform interpretation of Physical Optics scattering. Unlike point-sampling methods, our discrete implementation bins the profile in range while retaining a complex-valued representation that encodes sub-bin phase corrections, so the field can be evaluated accurately at frequencies well beyond the nominal mesh resolution. We also derive a corresponding time-domain formulation for the scattered transient response to analytic incident pulses. We further show that the classical second-derivative area law for far-field transient scattering extends into the near field as a two-term expression in the derivatives of an isochrone projected area, which reduces to the classical law as the additional induction term decays with range.
\end{abstract}


\IEEEpeerreviewmaketitle

\section{Introduction}

\IEEEPARstart{T}{he} accurate and efficient analysis of high-frequency wideband scattering by large, geometrically complex objects is required in numerous applications, among them target identification, synthetic aperture imaging, and radar cross-section (RCS) evaluation. The analysis often has to be repeated for many excitations and across a wide frequency band, on a single workstation of limited storage. Numerically exact integral equation solvers, such as the method of moments (MoM) \cite{Peterson1998}, combined with fast field evaluation algorithms \cite{Engheta1992, Brick2010, Chernokozhin2016}, can treat targets of a few dozen wavelengths in size on such a platform. For larger targets, and without access to high-performance computing resources, approximate asymptotic methods remain the practical alternative.

These divide into two families. The first is based on the geometrical optics (GO) approximation \cite{Deschamps1972}, extended to account for edge diffraction by the geometrical and the uniform theories of diffraction \cite{Keller1962, Kouyoumjian1974}. The second is based on the physical optics (PO) approximation \cite{MacDonald1912}, which by itself does not represent diffraction and is therefore augmented with the physical theory of diffraction \cite{Ufimtsev2007} or with equivalent edge currents \cite{Michaeli1984}. Both families map well onto parallel hardware \cite{Tao2010, Wei2012}.

In ray-based methods the incident field is carried by a bundle of rays traced through multiple reflections, and the scattered amplitude follows from the ray density on exit, so multiple reflections and shadowing are handled by the tracing itself at a frequency-independent cost. Two drawbacks follow: the ray density diverges at caustics, which shooting and bouncing rays (SBR) \cite{Ling1989} and ray-density normalization \cite{Didascalou2000} were introduced to control, and a ray field carries only the specular part of the scattered field.

PO proceeds differently. The surface sources are obtained from the incident field through a local tangent plane approximation, are assigned only to the lit part of the scatterer, and the scattered field follows from a radiation integral over them. The resulting current distribution is continuous over the surface, but on its own it agrees with numerically exact methods only for convex geometries at high frequencies. For nonconvex bodies, multiple reflections are reintroduced by the iterative PO (IPO) method \cite{Obelleiro1995}, in which the sources obtained at one iteration are radiated back onto the surface, where the resulting field induces the sources of the next order, so that the induced currents accumulate successive bounces. Every iteration then requires field integrals from the surface onto itself, at an $\mathcal{O}(N^2)$ cost, together with the mutual visibility between surface points. Techniques for accelerating these evaluations \cite{Burkholder2001, Burkholder2009, BurkholderLundin2005, BoagMichielssen2004, Roudstein2015, Gershenzon2018, Bourlier2019} have been developed, significantly reducing runtime and memory requirements.

All of these schemes accelerate the \textit{evaluation} of the PO integral; the quantity evaluated remains frequency-dependent. Discretization, subdomain count, runtime and memory continue to scale with the electrical size of the target, so widening the band or pushing the upper frequency still carries a proportional cost. Frequency-domain solvers must re-evaluate the oscillatory kernel $e^{-jkL}$ across a spectral grid dense enough to satisfy the Nyquist criterion of the target's impulse response. Adaptive-mesh formulations use the phase gradient as a refinement criterion to mitigate oversampling \cite{Burkholder2005, Song2023, Guo2026}, but they remain inside this loop, and they apply the criterion to reduce the cost of single-bounce evaluation; Section \ref{subsec:phase_space_amr} uses a related criterion for a different purpose, to resolve the gradient mismatch intrinsic to double-bounce interactions. Machine learning has also been applied to the problem, to extrapolate the RCS in frequency \cite{Han2025ML} and to predict wideband scattering directly from a surface mesh \cite{Jiang2026GTC}.

The time domain avoids this spectral grid altogether. Kennaugh and co-workers \cite{Kennaugh1958, Kennaugh1965} established that the far-field transient response is proportional to the second spatial derivative of the illuminated cross-sectional area, and Sun and Rusch \cite{SunRusch1994} carried the PO approximation itself into the time domain. Bölükbaş and Ergin \cite{Bolukbas2004Radon} then showed that the time-domain PO integral over a linear triangle admits a closed form when read as a Radon transform (RT), the surface integral collapsing onto the intersection of the facet with a moving plane. That interpretation has since been extended in two directions. Geometrically, it has been carried to NURBS surfaces \cite{SerimErgin2008}, arbitrary polygons \cite{ZhouCui2012}, and quadratic \cite{Aktepe2020} and higher-order \cite{Aktepe2023} triangles, and folded into an improved TDPO in which the closed form replaces numerical quadrature \cite{Guan2011}. In the field regime, Ülkü and Ergin \cite{UlkuErgin2010} treated all four source/observer configurations for acoustic scattering from soft and hard surfaces, the moving plane becoming a curved quadric surface whenever either end lies in the near field. Karaca and Ergin \cite{Karaca2015} derived the corresponding vector expressions for a near dipole source observed at infinity, while the near field has been reached differently, by replacing the single far-field expansion center of the Green's function with a separate expansion about each facet: Zhou and Cui \cite{ZhouCui2015} modified the phase in this way, and Chen and Tong \cite{ChenTong2019} carried the correction through to the amplitude as well, so that both remain accurate at short range. Curved iso-delay and near-field configurations are therefore already well established in this line of work.

Two limitations motivate the present work. The first concerns how the delay axis is resolved. The RT evaluates an instantaneous line integral, exact wherever it samples, but it samples only on the time grid, and whatever lies between consecutive samples is lost. This is not confined to the pathological case in which an entire scatterer falls between two samples, for which Aktepe and Ülkü \cite{Aktepe2021} propose an area-preserving rectangular substitute. It applies wherever a facet's delay span is comparable to the time step, and at the ends of a target's delay support, where the surface remaining beyond the last sample that registered anything is absent from the result. The error is therefore governed by the temporal step and the observation direction, and refining the mesh does not remove it.

The second limitation is the scattering order. Every formulation above, whether it yields a closed form or a numerical evaluation, addresses single-bounce scattering. Multiple reflections are treated instead by Iterative PO \cite{Obelleiro1995}, and whose time-domain counterpart computes the coupled surface currents numerically at each step \cite{li2011tdipo}, more recently with ray tracing used to restrict the coupling to local regions and with GPU parallelization \cite{Sun2023TDLIPO}. To our knowledge the contour-based slicing has not been extended to multiple bounces: no level-set representation of a double-bounce field, analogous to the single-bounce RT constructions above, has been reported.

This paper takes a different approach, aimed at how the result of the PO integral is represented along the delay axis rather than at how the integral is evaluated over a facet. The Co-Area formula \cite{Federer1969} expresses a surface integral as an integral over the level sets of an arbitrary scalar field, weighted by the local gradient magnitude $|\nabla_S L|$. Applied to the total delay, it makes the surface phase gradient the governing quantity: the profile is assembled from local, pointwise slicing of the phase, with no reference to the global form of the level set, so near- and far-field configurations and first- and second-order interactions are instances of one construction. The resulting frequency-independent envelopes are termed Geometric Range Profiles (GRPs), and their assembly is analytic throughout, requiring no numerical quadrature at either order.

The per-facet slicing itself is not new: under plane-wave illumination observed in the far field it reduces to the same triangle function as the Radon-transform closed forms, and in the near field the present work shares its local-linearity treatment with locally expanded time-domain formulations \cite{ChenTong2019}, while conic-intersection formulations \cite{UlkuErgin2010, Karaca2015} instead retain the exact curved iso-delay surface at the cost of a separate derivation per configuration. The difference lies in what is done with each slice. Rather than evaluating the field where an iso-delay surface cuts a facet at one instant, the proposed Co-Area Phase Shifting (CAPS) algorithm integrates the facet's whole intersection with a bin of finite width, and represents that contribution by two numbers instead of by its exact phase integral: a real magnitude and the exact sub-bin position of its phase center of mass, carried as a complex residual. Contributions from all facets meeting a bin are then summed coherently into one complex number.

An exact phase integral over a slice cannot be re-evaluated at a new wavenumber without being recomputed, whereas a magnitude and a position can simply be rescaled. The price is that intra-bin phase cancellation is no longer carried exactly and must be restored by a shape-factor correction, which bounds the usable bin width; the return is that one extraction at a reference frequency serves an entire band, and that a first-order profile can be translated intact, by a phase multiplication alone, when it becomes the illuminating field memory for a second bounce. This translatability, rather than the slicing itself, is what makes the second order tractable: the field arriving at a point of $S_2$ is a distributed profile rather than an analytic source, and a time-sampled history offers no sub-sample information with which to shift it to a neighboring point.

Accuracy therefore remains governed by the bin width rather than being exact, but the sub-bin representation loosens that dependence considerably: Section \ref{subsec:results_sphere_radon} matches the closed-form Radon-transform solution at range-sampling intervals roughly two orders of magnitude coarser than a point-sampling time-domain solver requires. The same construction also recasts classical transient scattering theory in the near field. Near-field time-domain PO expressions are themselves well established \cite{UlkuErgin2010, Karaca2015, ChenTong2019}; what the Co-Area formulation adds is that they can be written, as in the classical far-field result, as derivatives of a projected area (here an isochrone projected area on curved iso-delay contours rather than flat cross sections). An induction term proportional to the first area derivative accompanies the classical radiation term, and the Kennaugh--Cosgriff second-derivative law is recovered exactly in the far-field limit, where the induction term decays as $1/R_o^2$ (Appendix \ref{app:km_generalization}). Throughout, the continuous mesh evaluation captures area-jump edge singularities directly.

A further consequence, developed in Section \ref{sec:time_domain}, is that the discrete GRP arrays translate directly into exact time-domain convolution kernels. Each entry carries a real magnitude together with a sub-bin range residual embedded as complex phase, so the transient response follows either by applying that residual as an exact fractional delay to the continuous, analytically differentiable input signal, or, for practical passband simulation, by convolving the complex array against the signal's baseband envelope. Both routes avoid the naive discrete convolution that would collapse all sub-bin timing information onto the coarse grid.

The continuous formulation of both orders places no restriction on the shape of the scatterer. The discrete second-order engine does. Translating the field memory between neighboring points on $S_2$ relies on a single dominant arrival direction, which the argument of Section \ref{subsec:discrete_2nd_order_bottleneck} establishes only for a convex illuminating segment free of caustics. The double-bounce results reported here are therefore restricted to piecewise-convex bodies interacting through a single pair of distinct surfaces, and the present engine should not be applied to generally concave geometries, where focusing of the intermediate field would invalidate that assumption. Extending it to further interactions between piecewise-convex surfaces is a more modest step, requiring segmentation of the target and bookkeeping of several contributors rather than a different treatment of the intermediate field. Finally, as a physical optics formulation the method evaluates the PO integral itself, including the area-jump behavior of its edge truncation; the fringe-current correction of the physical theory of diffraction, incorporated in some time-domain solvers through equivalent edge currents \cite{li2011tdipo, ChenTong2019}, is an additive term outside the present scope.

The remainder of this paper is organized as follows. Section \ref{sec:problem_formulation} establishes the problem formulation and isolates the spectral bottleneck. Section \ref{sec:first_order} derives the first-order GRPs from the Co-Area formula and phase level sets. Section \ref{sec:second_order} extends the construction to second-order interactions, introducing the dual phase-space Adaptive Mesh Refinement that controls the incoming-outgoing gradient mismatch. Section \ref{sec:extrapolation} develops the wideband frequency extrapolation common to both orders. Section \ref{sec:time_domain} reconstructs the transient response from the same complex arrays. Section \ref{sec:results} presents the numerical validation: a sphere against Radon-transform PO, a near-field flat plate against TDPO, and two double-bounce cases, a dihedral and a half-ellipsoid on a plate, against fully converged IPO and asymptotic references. Finally, Section \ref{sec:conclusion} concludes the paper.

\section{Problem Formulation}
\label{sec:problem_formulation}

Let $S$ be a piecewise smooth impenetrable surface, possibly made of several disjoint bodies, such as the union of two closed surfaces $S_1$ and $S_2$. An incident scalar wave field $\varphi_{\text{inc}}$ illuminates $S$ and gives rise to a scattered field $\varphi_{\text{scat}}$, so that the total field $\varphi_{\text{tot}} = \varphi_{\text{inc}} + \varphi_{\text{scat}}$ obeys the Helmholtz equation $(\nabla^2 + k^2)\varphi_{\text{tot}} = 0$ outside $S$, with $k$ the wavenumber and a time-harmonic dependence $e^{j \omega t}$ assumed and suppressed. A Neumann boundary condition is imposed on $S$, and the radii of curvature of the geometry are taken to be large compared with the wavelength, so that the PO approximation evaluates the scattered field with sufficient accuracy.

Following the Iterative PO (IPO) approach \cite{Gershenzon2018}, the induced surface field is approximated as a superposition of successive reflection orders, $\varphi(\mathbf{r}) \approx \sum_{m=1}^{\infty} \varphi_m(\mathbf{r})$, whose first term is the PO source term
\begin{equation}
\varphi_1(\mathbf{r}) = 
\begin{cases} 
2 \varphi_{\text{inc}}(\mathbf{r}) & \text{lit region} \\ 
0 & \text{shadow region} 
\end{cases}
\label{eq:po_first_order}
\end{equation}

Higher-order surface fields ($m > 1$) are induced recursively. With the Green's function $G(\mathbf{r}, \mathbf{r}') = e^{-j k R}/{4 \pi R}$, $R \triangleq |\mathbf{r} - \mathbf{r}'|$, the $m$-th order surface field $\varphi_{m}$ radiates the field observed at a point $\mathbf{r}$:
\begin{equation}
\varphi^{(m)}(\mathbf{r}) = \iint_{S} \varphi_{m}(\mathbf{r}'') \partial_{n''} G(\mathbf{r}, \mathbf{r}'') V(\mathbf{r}, \mathbf{r}'') \, dS''
\label{eq:m_radiated_field}
\end{equation}
where $\partial_{n''}$ denotes the normal derivative with respect to the outward-pointing unit normal vector $\hat{\mathbf{n}}''$ at the source point $\mathbf{r}''$, and $V(\mathbf{r}, \mathbf{r}'')$ is the binary visibility function. 

Evaluated at a second surface point $\mathbf{r}' \in S$, this field is the internal incident field arriving from the previous bounce. For a rigid target, the surface field it induces is exactly twice that internal field:
\begin{equation}
\varphi_m(\mathbf{r}') = 2 \varphi^{(m-1)}(\mathbf{r}')
\label{eq:recursive_relation}
\end{equation}

Substituting the radiation integral \eqref{eq:m_radiated_field} into \eqref{eq:recursive_relation} gives the recursion for the internal surface field:
\begin{equation}
\varphi_{m}(\mathbf{r}') = 2 \iint_{S} \varphi_{m-1}(\mathbf{r}'') \partial_{n''} G(\mathbf{r}', \mathbf{r}'') V(\mathbf{r}', \mathbf{r}'') \, dS''
\label{eq:recursive_surface_field}
\end{equation}

The scattered field reaching a receiver at $\mathbf{r}$ is the sum of all radiation orders: $\varphi_{\text{s}}(\mathbf{r}) = \sum_{m=1}^{\infty} \varphi^{(m)}(\mathbf{r})$.

\subsection{The Spectral Bottleneck}
Equation \eqref{eq:recursive_surface_field} describes the multi-bounce topology compactly, but for broadband applications it exposes a spectral bottleneck. The free-space Green's function carries the highly oscillatory phase kernel $e^{-jk|\mathbf{r}-\mathbf{r}'|}$, so extracting a wideband response traditionally requires re-evaluating these dense spatial integrals over a frequency grid fine enough to satisfy the Nyquist criterion. The following section applies Co-Area integration to this spatial-spectral coupling, projecting the surface interactions analytically onto exact one-dimensional Baseband Geometric Range Profiles (BGRPs).

\section{First-Order Geometric Range Profile}
\label{sec:first_order}

To construct the BGRPs, we begin with the first-order scattered field under the PO approximation. Isolating the rapidly oscillating phase exponential, it can be written in the general form
\begin{equation}
\varphi^{(1)}(\mathbf{r}) = \iint_{S} g(\mathbf{r}, \mathbf{r}', k) e^{-j k L(\mathbf{r}')} \, dS'
\label{eq:first_order_general}
\end{equation}
where $L(\mathbf{r}') = L_{\text{inc}}(\mathbf{r}') + L_{\text{out}}(\mathbf{r}')$ is the total bistatic spatial delay. Both legs admit a near-field and a far-field form, depending on whether the source and the observer lie at a finite distance or effectively at infinity:
\begin{equation}
L_{\text{inc}}(\mathbf{r}') \!=\! \begin{cases} |\mathbf{r}'\!-\!\mathbf{r}_s| \\ \hat{\mathbf{k}}_{\text{inc}} \!\cdot\! \mathbf{r}' \end{cases}\!\!,\ \
L_{\text{out}}(\mathbf{r}') \!=\! \begin{cases} |\mathbf{r}\!-\!\mathbf{r}'| & \!\!\text{near field} \\ -\hat{\mathbf{k}}_{\text{out}} \!\cdot\! \mathbf{r}' & \!\!\text{far field} \end{cases}
\label{eq:l_cases}
\end{equation}
(point source and near-field observer on top, plane wave and far-field observer below), with $\hat{\mathbf{k}}_{\text{inc}}$, $\hat{\mathbf{k}}_{\text{out}}$ the fixed incidence/observation directions. The unit vector $\hat{\mathbf{R}}(\mathbf{r}')$ used below follows the same split: the direction from $\mathbf{r}'$ toward $\mathbf{r}$ in the near field, or the fixed direction $\hat{\mathbf{k}}_{\text{out}}$ in the far field. The incident field follows the same case distinction, either a spherical wave from a point source or an ideal plane wave,
\begin{equation}
\varphi_{\text{inc}}(\mathbf{r}') = \begin{cases} \dfrac{A_0}{4\pi|\mathbf{r}'-\mathbf{r}_s|}\,e^{-jk|\mathbf{r}'-\mathbf{r}_s|} \\[4pt] A_0\, e^{-jk(\hat{\mathbf{k}}_{\text{inc}}\cdot\mathbf{r}')} \end{cases}
\label{eq:phi_inc_cases}
\end{equation}
and this near/far-field choice is used consistently throughout the paper.

The complex amplitude kernel $g(\mathbf{r},\mathbf{r}',k)$ comes from the normal derivative of the free-space Green's function, which separates into a near-field term decaying as $1/|\mathbf{r}-\mathbf{r}'|^2$ and a far-field term decaying as $1/|\mathbf{r}-\mathbf{r}'|$. Factoring out the oscillatory phase $e^{-jkL(\mathbf{r}')}$ isolates a purely geometric \textit{core kernel},
\begin{equation}
\mathcal{G}^{(1)}(\mathbf{r},\mathbf{r}') = \frac{\hat{\mathbf{R}}(\mathbf{r}')\cdot\hat{\mathbf{n}}(\mathbf{r}')}{4\pi|\mathbf{r}-\mathbf{r}'|^2}, \qquad
\mathcal{G}^{(2)}(\mathbf{r},\mathbf{r}') = \frac{\hat{\mathbf{R}}(\mathbf{r}')\cdot\hat{\mathbf{n}}(\mathbf{r}')}{4\pi|\mathbf{r}-\mathbf{r}'|},
\label{eq:core_kernel_def}
\end{equation}
At first order the source term is the known closed-form incident field, so its magnitude is absorbed into the kernel along with the joint visibility $V(\mathbf{r}')=V_{\text{inc}}(\mathbf{r}_s,\mathbf{r}')V_{\text{s}}(\mathbf{r},\mathbf{r}')$:
\begin{equation}
\begin{split}
g(\mathbf{r}, \mathbf{r}', k) &= g^{(1)}(\mathbf{r}, \mathbf{r}') + j k g^{(2)}(\mathbf{r}, \mathbf{r}'), \\
g^{(i)} &= 2\,|\varphi_{\text{inc}}(\mathbf{r}')|\, V(\mathbf{r}')\, \mathcal{G}^{(i)}(\mathbf{r},\mathbf{r}')
\end{split}
\label{eq:kernel_split}
\end{equation}

\subsection{Continuous GRP Extraction via the Co-Area Formula}
\label{subsec:continuous_extraction}

Rather than integrating over the surface again at every frequency of a dense band, we apply the Co-Area formula \cite{Federer1969} as a geometric change of variables that turns the surface integral into an integral over iso-phase level sets. The level set at a spatial delay $l$ is $\Gamma(l) = \{ \mathbf{r}' \in S \mid L(\mathbf{r}') = l \}$.

The Co-Area formula states that for a scalar partitioning function $f$ with generalized Jacobian $Jf$,
\begin{equation}
\iint_{S} F(\mathbf{r}')\, Jf(\mathbf{r}') \, dS' = \int \left( \oint_{f^{-1}(y)} F(\mathbf{r}') \, d\gamma \right) dy
\label{eq:coarea_general_jacobian}
\end{equation}
where $d\gamma$ is the differential arc length along a level set. Here the partitioning function is the total delay restricted to the surface, $f = L|_S$, whose level sets are the iso-phase contours $\Gamma(l)$. Its Jacobian follows from the surface-gradient operator, defined for an arbitrary scalar field $\Psi$ on $S$ as the projection of the full 3D gradient onto the local tangent plane,
\begin{equation}
\nabla_S \Psi(\mathbf{r}') \triangleq \nabla \Psi(\mathbf{r}') - \left( \nabla \Psi(\mathbf{r}') \cdot \hat{\mathbf{n}}(\mathbf{r}') \right) \hat{\mathbf{n}}(\mathbf{r}')
\label{eq:surface_gradient}
\end{equation}
which is the rate of surface phase progression at a point $\mathbf{r}'$ on $S$. The Jacobian then reduces to $J(L|_S) = |\nabla_S L(\mathbf{r}')|$, or equivalently
\begin{equation}
dS' = \frac{d\gamma \, dl}{|\nabla_S L(\mathbf{r}')|}
\label{eq:differential_area_relation}
\end{equation}

Substituting $y=l$ and $F = g^{(i)}(\mathbf{r},\mathbf{r}')\,e^{-jkL(\mathbf{r}')} / |\nabla_S L(\mathbf{r}')|$ into \eqref{eq:coarea_general_jacobian} returns, on the left, the $i$-th component of the first-order integral \eqref{eq:first_order_general}. On the right, $L(\mathbf{r}') = l$ is constant along $\Gamma(l)$, so the oscillatory factor leaves the contour integral and the wavenumber decouples from the surface geometry. What remains inside is purely geometric, and defines the two continuous BGRPs:
\begin{equation}
h_B^{(i)}(l) = \oint_{\Gamma(l)} \frac{g^{(i)}(\mathbf{r}, \mathbf{r}')}{|\nabla_S L(\mathbf{r}')|} \, d\gamma, \quad i \in \{1,2\}
\label{eq:continuous_bgrps}
\end{equation}
in terms of which the total first-order field becomes a Fourier-type transform of a linear, time-invariant response:
\begin{equation}
\varphi^{(1)}(\mathbf{r}, k) = \int_{l_{min}}^{l_{max}} \left[ h_B^{(1)}(l) + j k h_B^{(2)}(l) \right] e^{-j k l} \, dl
\label{eq:continuous_lti}
\end{equation}
Applying the inverse Fourier transform and mapping range to time ($l = ct$, $jk \rightarrow \frac{1}{c}\frac{d}{dt}$) recovers the continuous time-domain transient impulse response:
\begin{equation}
\tilde{h}^{\text{I}}(t) = c\, h_B^{(1)}(ct) + \frac{d}{dt} h_B^{(2)}(ct)
\label{eq:temporal_h1}
\end{equation}

\subsection{Discrete Implementation and Analytical Slicing}
\label{subsec:discrete_implementation}

To implement this continuous formulation, the surface $S$ is approximated by a discrete triangulated mesh $S_D = \bigcup_{t \in \mathcal{T}} t$ with $\mathcal{T}$ the set of triangles comprising the mesh, and the delay axis is binned into equally spaced bins $l[n] = (n+0.5)dl$ of width $dl$. The discrete bin value is, by definition, the continuous profile integrated across the bin width. Inverting the Co-Area relation \eqref{eq:differential_area_relation} returns an ordinary surface integral; restricting it to the discrete mesh and decomposing over the triangles $t \in T_n$ that intersect the strip gives
\begin{equation}
\begin{split}
h^{(i)}[n] &= \frac{1}{dl}\int_{l[n]-\frac{dl}{2}}^{l[n]+\frac{dl}{2}} \left[ \oint_{\Gamma(l)} \frac{g^{(i)}(\mathbf{r}, \mathbf{r}')}{|\nabla_S L(\mathbf{r}')|} \, d\gamma \right] e^{-jkl} \, dl \\
&\approx \frac{1}{dl}\sum_{t \in T_n} \iint_{\Delta S_{n,t}} g^{(i)}(\mathbf{r}, \mathbf{r}') e^{-jkL(\mathbf{r}')}\,dS'
\end{split}
\label{eq:nested_bin_integration}
\end{equation}
where, for a delay field $L$ partitioning a discrete surface $S_D$ into bins centered at $l[n]$, the strip is $\Delta S_n^D = \{\mathbf{r}' \in S_D \mid |L(\mathbf{r}') - l[n]| \le dl/2\}$, $T_n = \{t \in \mathcal{T} \mid t \cap \Delta S_n^D \neq \emptyset\}$ are the triangles meeting it, and $\Delta S_{n,t} = t \cap \Delta S_n^D$ is the slice of triangle $t$ within it.

Evaluating the slice integral requires two approximations. The kernel $g^{(i)}(\mathbf{r},\mathbf{r}')$ varies smoothly over a facet, so it is evaluated once at the triangle's centroid $\mathbf{r}'_c$ and pulled outside the integral, giving the discrete amplitude coefficient $g^{(i)}[t] = g^{(i)}(\mathbf{r},\mathbf{r}'_c)\,A[t]$ (with $A[t]$ the triangle's total area). The rapidly oscillating phase $e^{-jkL(\mathbf{r}')}$ cannot be treated the same way; its contribution over the slice defines an effective phase center of mass (COM) $l_{\text{COM}}$,
\begin{equation}
\iint_{\Delta S_{n,t}} e^{-jkL(\mathbf{r}')}\,dS' \triangleq A_{\text{pol}}[n,t]\,e^{-jkl_{\text{COM}}},
\label{eq:exact_slice_phase}
\end{equation}
with $A_{\text{pol}}[n,t]$ the slice's exact physical area. 

Evaluating \eqref{eq:exact_slice_phase} directly poses two difficulties. The slice area $A_{\text{pol}}[n,t]$ requires the boundary vertices of the clipped polygon $\Delta S_{n,t}$, and the phase COM requires resolving how that area is distributed in delay across the bin. Assigning the phase to the bin center $l[n]$ causes phase errors at high frequencies, especially near stationary-phase regions where the strip is wide and the slice departs substantially from the center, and it still leaves the polygon area to be computed.

Both difficulties are resolved by assuming that the incident wavefront is locally planar over a facet, so that the spatial delay varies linearly across it. The area and the phase COM are then the zeroth and first moments of a hat function, obtained by analytic integration. Sorting a triangle's vertex delays as $l_{\text{min}} \le l_{\text{mid}} \le l_{\text{max}}$, the hat function is
\begin{equation}
w(l) = 
\begin{cases} 
C_{\text{rise}}(l - l_{\text{min}}) & l_{\text{min}} \le l < l_{\text{mid}} \\ 
C_{\text{fall}}(l_{\text{max}} - l) & l_{\text{mid}} \le l \le l_{\text{max}} 
\end{cases}
\label{eq:hat_function}
\end{equation} with $C_{\text{rise}}, C_{\text{fall}}$ chosen so $\int w\,dl = 1$. Geometrically this is the same triangular density that underlies the Radon Transform (RT) interpretation of PO scattering introduced by Bölükbaş et al.~\cite{Bolukbas2004Radon}, but it is used differently here: their expression computes the contribution of a single triangle at a time sample on a fixed grid, whereas ours locates a slice's phase COM continuously within a bin, to be summed into a weighted phase of the level-set strip rather than restricted to rigid range steps. Applying the phase linearity per facet also admits a globally spherical wavefront, as established in \cite{ChenTong2019}.

For a triangle intersecting the bin over $[l_a,l_b]$, its fractional area (the zeroth moment of $w(l)$) and the corresponding path-length-weighted first moment follow by analytic integration:
\begin{equation}
M_{\text{0}}[n,t] = \int_{l_a}^{l_b} w(l)\,dl, \qquad M_{\text{1}}[n,t] = \int_{l_a}^{l_b} l\, w(l)\,dl.
\label{eq:s_m_total}
\end{equation}
The slice's phase COM and residual follow immediately,
\begin{equation}
l_{\text{COM}} = \frac{M_{\text{1}}[n,t]}{M_{\text{0}}[n,t]}, \qquad \Delta l_n^t = l_{\text{COM}} - l[n],
\label{eq:slice_residual}
\end{equation}
and this is the \textit{Slice Residual} $\Delta l_n^t$, which lets many triangles be combined coherently into one bin in a complex representation. To our knowledge, encoding the sub-bin position as a complex residual in a shared bin, rather than as an exact independent closed form per facet, is new, and the rest of this paper builds on it.

With the residual in hand, the discrete BGRP is assembled as one complex number per bin. We refer to this discrete realization, the binned profile together with the phase residual that carries sub-bin position through every subsequent operation, as the CAPS algorithm, to distinguish it from the continuous Co-Area formulation of Sections \ref{subsec:continuous_extraction} and \ref{subsec:continuous_2nd_order}:
\begin{equation}
h_B^{(i)}[n] = \frac{1}{dl} \sum_{t \in T_n} g^{(i)}[t]\, M_{\text{0}}[n,t]\, e^{-j k \Delta l_n^t}.
\label{eq:discrete_baseband}
\end{equation}
This carries a practical consequence. Because \eqref{eq:discrete_baseband} integrates each triangle's exact area rather than sampling it at discrete instants, it does not suffer the ``Under-Sampled Reflection'' failure mode identified in \cite{Aktepe2021}, as demonstrated in Section \ref{subsec:results_sphere_radon}.

The \textit{Strip Residual}, $\Delta \bar{l}_n = -\frac{1}{k}\angle(h_B^{(i)}[n])$, is the phase of this sum: it anchors the aggregated field at its true sub-bin position, with accuracy bounded by the local-linearity assumption above rather than by the bin width $dl$, whose own limitation is described later.
Both profiles are extracted once, at the reference wavenumber $k$, and carry no frequency dependence beyond the phase residuals; their evaluation at an arbitrary $k'$ is explained in Section \ref{sec:extrapolation}, together with the second-order case.

\section{Second-Order Geometric Range Profile}
\label{sec:second_order}

This section extends the first-order derivation of Section \ref{sec:first_order} to the double-bounce case. We first derive the continuous second-order profile (Section \ref{subsec:continuous_2nd_order}), and then construct the discrete algorithm that evaluates it efficiently, reusing the first-order GRPs for the double-bounce profiles assembly through the phase corrections described below.

In a double-bounce mechanism, the secondary surface $S_2$ is illuminated by the near field radiated by the primary surface $S_1$. Setting $m=2$ and substituting the internal incident field $\varphi^{(1)}(\mathbf{r}')$ of \eqref{eq:m_radiated_field} into the PO induced surface field \eqref{eq:recursive_relation} gives the secondary outgoing scattered field in the general form
\begin{equation}
\varphi^{(2)}(\mathbf{r}) = \iint_{S_2} \varphi^{(1)}(\mathbf{r}') g_2(\mathbf{r}, \mathbf{r}', k) e^{-j k L_2(\mathbf{r}')} \, dS_2
\label{eq:second_order_general}
\end{equation}
where $L_2(\mathbf{r}')$ is the outgoing spatial delay from $S_2$ to the observer $\mathbf{r}$, and $g_2(\mathbf{r}, \mathbf{r}', k)$ is the secondary complex amplitude kernel defined as:
\begin{equation}
\begin{split}
g_2(\mathbf{r}, \mathbf{r}', k) &= g_2^{(1)}(\mathbf{r}, \mathbf{r}') + j k g_2^{(2)}(\mathbf{r}, \mathbf{r}'), \\
g_2^{(i)} &= 2\,V(\mathbf{r},\mathbf{r}')\, \mathcal{G}^{(i)}(\mathbf{r},\mathbf{r}')
\end{split}
\label{eq:g2_split}
\end{equation}
Here $V(\mathbf{r}, \mathbf{r}')$ is the target-to-receiver visibility function.

\subsection{Continuous Geometric Range Profiles}
\label{subsec:continuous_2nd_order}

As in Section \ref{sec:first_order}, we apply the Co-Area formula to \eqref{eq:second_order_general}, mapping $S_2$ onto the outgoing delay axis $l_2$ with level sets $\Gamma(l_2) = \{ \mathbf{r}' \in S_2 \mid L_2(\mathbf{r}') = l_2 \}$. Applying \eqref{eq:surface_gradient} to the outgoing delay $L_2$, the generalized Jacobian reduces to $J(L_2|_S) = |\nabla_S L_2(\mathbf{r}')|$, and \eqref{eq:coarea_general_jacobian} gives
\begin{equation}
\varphi^{(2)}(\mathbf{r}) = \int_{l_{2}} \left[ \oint_{\Gamma(l_2)} \frac{\varphi^{(1)}(\mathbf{r}') g_2(\mathbf{r}, \mathbf{r}', k)}{|\nabla_S L_2(\mathbf{r}')|} \, d\gamma \right] e^{-j k l_2} \, dl_2
\label{eq:u2_coarea_applied}
\end{equation}

The incident field $\varphi^{(1)}(\mathbf{r}')$ arriving at $S_2$ is reconstructed via the primary consolidated BGRP $\bar{h}_B^{\text{I}} = h_B^{(1)} + j k h_B^{(2)}$ along the intermediate delay axis $l_1$: $\varphi^{(1)}(\mathbf{r}') = \int \bar{h}_B^{\text{I}}(l_1, \mathbf{r}') e^{-j k l_1} \, dl_1$. Substituting this into \eqref{eq:u2_coarea_applied} yields a nested integral that depends on both $l_1$ and $l_2$.

To map this integral onto the total delay $l$, we change variables as $l_1 = l - l_2$. Exchanging the order of integration reduces the scattered field to a one-dimensional Fourier-type transform over the total delay bounds:
\begin{equation}
\begin{split}
\varphi^{(2)}(\mathbf{r}) = \int_{l_{min}}^{l_{max}} \Bigg[ \int_{l_{2}} \oint_{\Gamma(l_2)} &\frac{\bar{h}_B^{\text{I}}(l - l_2, \mathbf{r}') g_2(\mathbf{r}, \mathbf{r}', k)}{|\nabla_S L_2(\mathbf{r}')|} \, d\gamma \, dl_2 \Bigg] \\
&\times\, e^{-j k l} \, dl
\end{split}
\label{eq:u2_change_of_variables_compact}
\end{equation}

Expanding $\bar{h}_B^{\text{I}}$ and $g_2$, the second-order spatial impulse response partitions into four frequency-independent one-dimensional BGRPs:
\begin{equation}
h_B^{(i j)}(l) = \int_{l_{2}} \oint_{\Gamma(l_2)} \frac{h_B^{(i)}(l - l_2, \mathbf{r}') g_2^{(j)}(\mathbf{r}, \mathbf{r}')}{|\nabla_S L_2(\mathbf{r}')|} \, d\gamma \, dl_2
\label{eq:direct_bgsp_transition}
\end{equation}
for $i, j \in \{1, 2\}$. Grouping these real-valued profiles by their wavenumber coefficients yields the second-order scattered field representation:
\begin{equation}
\varphi^{(2)}(\mathbf{r}) = \int_{l_{min}}^{l_{max}} \left[ h_B^{(11)}(l) + j k \Sigma_B^{(1)}(l) - k^2 h_B^{(22)}(l) \right] e^{-j k l} \, dl
\label{eq:final_u2_1d}
\end{equation}
where $\Sigma_B^{(1)}(l) = h_B^{(21)}(l) + h_B^{(12)}(l)$. Applying the inverse Fourier transform with the range to time mapping recovers the continuous time-domain transient impulse response:
\begin{equation}
\tilde{h}^{\text{II}}(t) = c h_B^{(11)}(ct) + \frac{d}{dt} \Sigma_B^{(1)}(ct) + \frac{1}{c} \frac{d^2}{dt^2} h_B^{(22)}(ct)
\label{eq:temporal_h2}
\end{equation}

\subsection{Independent Partitioning and the Field Memory Translation}
\label{subsec:discrete_2nd_order_bottleneck}

A direct discretization of \eqref{eq:direct_bgsp_transition} would partition the delay axis by the \textit{total} path $L = L_1 + L_2$. Because $L_2(\mathbf{r}')$ varies independently of $L_1$ over $S_2$, the corresponding $L_1$ level sets on $S_1$ would need to be re-sliced from scratch for every integration point on $S_2$, which is computationally prohibitive.

CAPS avoids this by partitioning $S_1$ and $S_2$ independently, splitting the outgoing delay into bins $p[m] = (m+0.5)dl$. The exact but expensive target quantity is the discrete bin value $\hat{h}^{(ij)}[n]$, obtained by integrating $h^{(ij)}(l) = h_B^{(ij)}(l) e^{-j k l}$ across the master bin partitioning $l$. Substituting $p[m]$ into it, applying the Co-Area relation \eqref{eq:differential_area_relation} and discretizing the surface yields:
\begin{equation}
\begin{split}
\hat{h}^{(ij)}[n] &= \frac{1}{dl} \sum_{m} \int_{l[n]-dl/2}^{l[n]+dl/2} \Bigg[ \sum_{t \in T_m} \iint_{\Delta S_{m,t}} \\
&\qquad h_B^{(i)}(l - L_2(\mathbf{r}'), \mathbf{r}') \, g_2^{(j)}(\mathbf{r}, \mathbf{r}') \, dS_2' \Bigg] e^{-j k l} \, dl
\end{split}
\label{eq:h2_strip_decomposition}
\end{equation}
where the strip, its triangle set and its slices, $\Delta S_m^D$, $T_m$ and $\Delta S_{m,t}$, follow the definitions of Section \ref{subsec:discrete_implementation} with the outgoing delay $L_2$ partitioning the discretized secondary surface $S_{2,D}$ into bins centered at $p[m]$. The incoming field $h_B^{(i)}$ is a spatially varying profile arriving from $S_1$, so it cannot be pulled outside the integral, while the scattering kernel is evaluated as constant over a facet. To make \eqref{eq:h2_strip_decomposition} tractable, we evaluate the incoming profile across a slice as the profile at its COM, which yields
\begin{equation}
\hat{h}^{(ij)}[n] \approx \sum_{m}\sum_{t\in T_m} g^{(j)}[t]\, M_{\text{0}}[m,t]\, \hat{I}^{(m,t)}
\label{eq:h2_temporal_binning}
\end{equation}

\begin{equation}
\hat{I}^{(m,t)} = \frac{1}{dl}\int_{l[n]-dl/2}^{l[n]+dl/2} h_B^{(i)}\big(l - L_2(\mathbf{r}'_{\text{COM}}), \mathbf{r}'_{\text{COM}}\big)\, e^{-jkl}\,dl.
\label{eq:h2_temporal_integral_def}
\end{equation}
This integrates the first-order profile, referred to as the \textit{field memory}, at $\mathbf{r}'_{\text{COM}}$, which differs from the triangle's center $\mathbf{r}_{c,t}$, where the discrete profiles were evaluated.

Bridging that gap requires knowing how the field memory changes between neighboring points on $S_2$, and here an approximation is used. Physically, the memory aggregated at $\mathbf{r}_{c,t}$ is a superposition of contributions arriving from across $S_1$, so moving to a nearby point shifts each contribution by a different amount and the profile stretches non-uniformly. Evaluating that stretching exactly would require re-binning $S_1$ for every slice, the cost we are trying to avoid.

We therefore apply a high-frequency computational simplification. At high frequency the scattered field is dominated by geometrical-optics rays originating at localized scattering centers and edge truncations \cite{Pathak2021}, and such rays satisfy Fermat's principle within a regular neighborhood \cite{BornWolf1999}. The scope adopted here works in our favor: although the global target may be piecewise convex, a double-bounce interaction between two distinct surfaces restricts the observer's view of $S_1$ to a convex segment, whose reflected ray tubes strictly expand and therefore form no real caustics \cite{Pathak2021}. Absent a caustic, the region qualifies as a regular optical neighborhood, and the dominant contribution is anchored to a single ray of strict minimum phase, located at
\begin{equation}
\mathbf{r}_{\text{MP}}(\mathbf{r}_{c,t}) = \arg\min_{\mathbf{r}'' \in S_1} \left( L_{\text{inc}}(\mathbf{r}'') + |\mathbf{r}_{c,t} - \mathbf{r}''| \right).
\label{eq:mp_point_definition}
\end{equation}

Rather than stretching, we treat the \textit{entire} profile as translating rigidly along this dominant direction of arrival, which reduces the displacement to a scalar projection,
\begin{equation}
\delta^{(m,t)} = \hat{\mathbf{r}}_{\text{MP}}(\mathbf{r}_{c,t}) \cdot \left( \mathbf{r}'_{\text{COM}} - \mathbf{r}_{c,t} \right)
\label{eq:mp_spatial_shift}
\end{equation}
\begin{equation}
\begin{split}
& \hat{\mathbf{r}}_{\text{MP}}(\mathbf{r}_{c,t}) \triangleq (\mathbf{r}_{c,t} - \mathbf{r}_{\text{MP}}(\mathbf{r}_{c,t}))/{L_{\text{MP}}(\mathbf{r}_{c,t})} \\
& L_{\text{MP}}(\mathbf{r}) = |\mathbf{r} - \mathbf{r}_{\text{MP}}(\mathbf{r})|
\end{split}
\label{eq:mp_direction}
\end{equation}
so $h_B^{(i)}(l, \mathbf{r}'_{\text{COM}}) \approx h_B^{(i)}(l - \delta^{(m,t)}, \mathbf{r}_{c,t})$. Substituting into \eqref{eq:h2_temporal_integral_def} yields:
\begin{equation}
\hat{I}^{(m,t)} \approx \frac{1}{dl}\int_{l[n]-dl/2}^{l[n]+dl/2} h_B^{(i)}\big(l - L_2(\mathbf{r}'_{\text{COM}}) - \delta^{(m,t)}, \mathbf{r}_{c,t}\big)\, e^{-jkl}\,dl,
\label{eq:h2_temporal_integral_approx}
\end{equation}
which turns $\hat{I}^{(m,t)}$ into a re-binning of the profile already computed at $\mathbf{r}_{c,t}$, shifted by the delay $L_2(\mathbf{r}'_{\text{COM}}) + \delta^{(m,t)}$.

The direct evaluation of the profile shifted by an arbitrary continuous delay $\Delta$ is a binning of the continuous profile across $l_1$ path bins indexed here by $\eta$,
\begin{equation}
\begin{split}
h_B^{(i)}[\eta, \Delta] \Big|_{\mathbf{r}_{c,t}} \triangleq \Bigg[ \frac{1}{dl}\int_{l[\eta]-dl/2}^{l[\eta]+dl/2} &h_B^{(i)}(l'-\Delta, \mathbf{r}_{c,t})\, e^{-jkl'}\,dl' \Bigg] \\
&\times\, e^{jkl[\eta]},
\end{split}
\label{eq:h2_generalized_baseband_binning}
\end{equation}
which is exact but expensive. CAPS instead defines a purely algebraic alternative to \eqref{eq:h2_generalized_baseband_binning}:
\begin{equation}
\check{h}_B^{(i)}[\eta, \Delta] \Big|_{\mathbf{r}_{c,t}} \triangleq h_B^{(i)}[\eta, 0] \Big|_{\mathbf{r}_{c,t}} \cdot e^{-jk\Delta}.
\label{eq:h2_discrete_projection}
\end{equation}
Taken as is, multiplying a physical real-valued range-profile sample by $e^{-jk\Delta}$ has no meaning in the range domain. This construction becomes exact because $h_B^{(i)}[\eta,\Delta]$ is not used by itself: it is re-modulated by the spatial carrier $e^{-jkl[\eta]}$ when the field is reconstructed via Fourier Transform. That combination gives $e^{-jk\Delta}\cdot e^{-jkl[\eta]} = e^{-jk(l[\eta]+\Delta)}$, the phase of a sample truly located at the shifted position $l[\eta]+\Delta$. In other words, \eqref{eq:h2_generalized_baseband_binning} and \eqref{eq:h2_discrete_projection} are algebraically identical once both are used within this reconstruction.

We now decompose the two delays entering \eqref{eq:h2_temporal_integral_approx} into an integer number of bins plus a sub-bin remainder: the outgoing delay as $L_2(\mathbf{r}'_{\text{COM}}) = (m+0.5)dl + \Delta l_{\text{out}}^{(m,t)}$, and the incoming translation as $\delta^{(m,t)} = K^{(m,t)}dl + \Delta l_{\text{in}}^{(m,t)}$. Changing the integration variable in \eqref{eq:h2_temporal_integral_approx} by the integer offset $l' = l-(m+K^{(m,t)})dl$ then maps the total path bin $n$ onto a parent bin $\eta = n-m-K^{(m,t)}$. Writing $\Delta = 0.5dl + \Delta l_{\text{out}}^{(m,t)} + \Delta l_{\text{in}}^{(m,t)}$ and using \eqref{eq:h2_generalized_baseband_binning} yields:
\begin{equation}
\hat{I}^{(m,t)} = h_B^{(i)}[\eta, \Delta] \Big|_{\mathbf{r}_{c,t}} \; e^{-jk\,l[n]},
\label{eq:h2_exact_reindex}
\end{equation}
 Replacing the binning operator by \eqref{eq:h2_discrete_projection}, writing the parent sample in polar form, $h_B^{(i)}[\eta,0] = | h_B^{(i)}[\eta,0] | e^{-jk\Delta l_{\text{parent}}^{(\eta,t)}}$ and applying to \eqref{eq:h2_exact_reindex} gives the quantity CAPS actually computes:
\begin{equation}
\begin{split}
I^{(m,t)} ={}& \Big| h_B^{(i)}\big[\eta,0\big] \Big|_{\mathbf{r}_{c,t}} \Big|\, e^{-jk\left(\Delta l_{\text{parent}}^{(\eta,t)} + \Delta l_{\text{in}}^{(m,t)} + \Delta l_{\text{out}}^{(m,t)}\right)} \\
&\times\, e^{-jk(n+1)dl},
\end{split}
\label{eq:h2_resolved_assembly}
\end{equation}
where $\Delta l_{\text{parent}}^{(\eta,t)}$ is the phase the parent array sample already carries.

The three residuals inside \eqref{eq:h2_resolved_assembly} are the three geometric quantities that must be computed. We summarize them next, after addressing why a single raw triangle $t$ is not always fine enough to evaluate them safely.

\subsection{Dual Phase-Space Adaptive Mesh Refinement (AMR)}
\label{subsec:phase_space_amr}

The Co-Area slicing partitions $S_2$ strictly by the outgoing delay, and so guarantees that $L_2$ varies by at most $dl$ within a slice. It guarantees nothing about the incoming delay. Applying the surface-gradient operator \eqref{eq:surface_gradient} to the dominant path of \eqref{eq:mp_point_definition}, and using the rigid translation established above, the incoming phase progression across a facet is governed by the tangential projection of the arrival direction:
\begin{equation}
\nabla_S L_{\text{MP}} = \hat{\mathbf{r}}_{\text{MP}}(\mathbf{r}_{c,t}) - \left( \hat{\mathbf{r}}_{\text{MP}}(\mathbf{r}_{c,t}) \cdot \hat{\mathbf{n}}(\mathbf{r}_{c,t}) \right) \hat{\mathbf{n}}(\mathbf{r}_{c,t})
\label{eq:mp_surface_gradient}
\end{equation}
Physically, $\nabla_S L_2$ points along the surface toward the observer while $\nabla_S L_{\text{MP}}$ points away from the illuminating region of $S_1$, so the two are geometrically independent and their mismatch takes two forms. A \textit{magnitude} mismatch arises when they are parallel but $L_{\text{MP}}$ varies faster, so that a single $L_2$ slice spans many incoming contours across its width. A \textit{directional} (cross-gradient) mismatch arises when they point in different directions, so that the slice cuts obliquely across the incoming contours and $L_{\text{MP}}$ varies along its length. In either case, sampling the field memory once at the slice centroid aliases it.

\begin{figure}[htbp]
    \centering
    \includegraphics[width=\columnwidth]{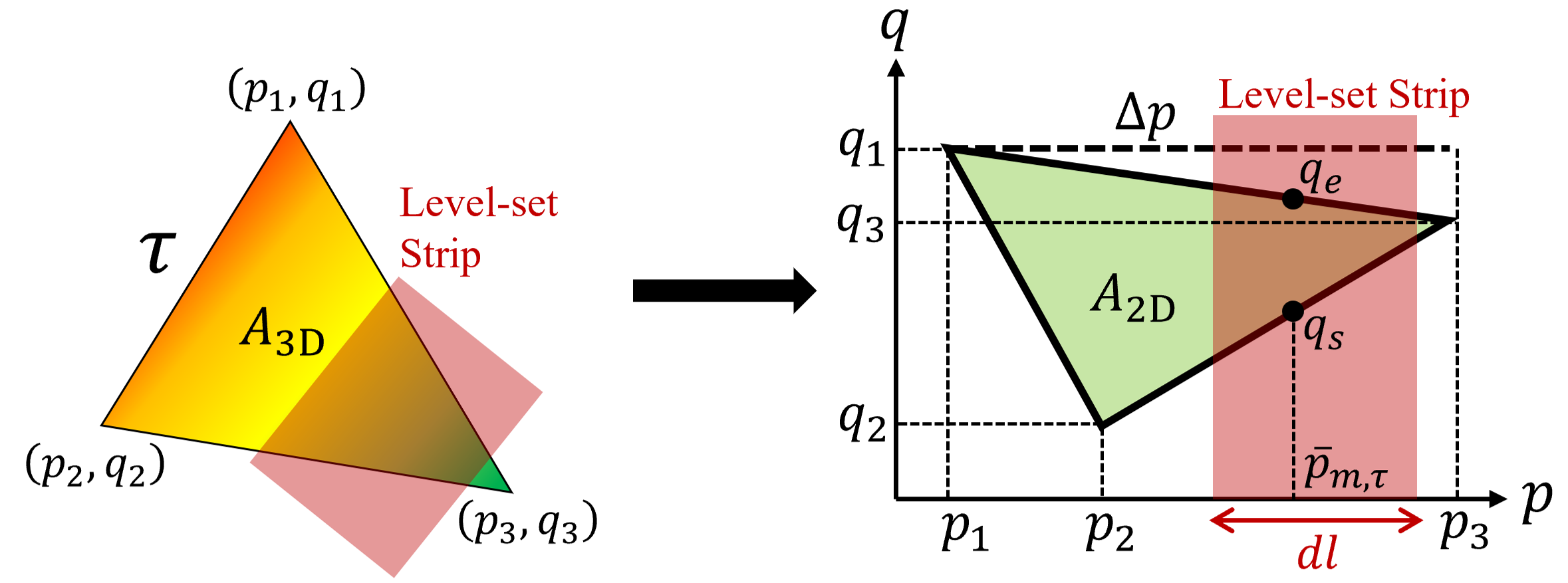}
    \caption{Affine transformation to the dual phase space.}
    \label{fig:pq_phase_space}
\end{figure}

To bound the incoming variation without globally refining the physical mesh, CAPS refines only where the mismatch demands it. Each parent triangle $t \in S_2$ is mapped, via its three vertices, into a 2D phase plane $(p,q)$ with $p = L_2(\mathbf{r}')$ and $q = L_{\text{MP}}(\mathbf{r}')$ (Fig. \ref{fig:pq_phase_space}). This affine map preserves barycentric ratios \cite{Zienkiewicz2025}, so the slicing problem can be solved equivalently in this plane.

Within this plane, the maximum possible spread of $q$ inside any outgoing bin of width $dl$ is bounded by
\begin{equation}
\Delta q_{\text{slice}} = \frac{2 A_{\text{2D}}}{\Delta p} + \max \left| \frac{dq}{dp} \right|_{\text{edges}} \cdot dl
\label{eq:total_q_variation}
\end{equation}
where $A_{\text{2D}}$ is the triangle's area in the $(p,q)$ plane and $\Delta p$ its vertex spread in $p$. Whenever $\Delta q_{\text{slice}} > dl$, the triangle is recursively quad-subdivided until every sub-triangle satisfies the bound, subject to a fixed limit on the recursion depth. We write $T_{\text{AMR},m}$ for the resulting set of sub-triangles intersecting the $m$-th outgoing strip.

Once the bound holds, both $p$ and $q$ vary by less than a bin across any slice, and both are affine over the plane. By Appendix \ref{app:phase_com_proof} the phase COM of a slice therefore coincides with its geometric centroid, and does so simultaneously for the outgoing and the incoming delay, so a single point serves for both. A slice $m$ intersecting $\tau \in T_{\text{AMR},m}$ is accordingly reduced to $(\bar{p}_{m,\tau}, \bar{q}_{m,\tau})$: the outgoing coordinate is the moment ratio $\bar{p}_{m,\tau} = M_1[m,\tau]/M_0[m,\tau]$ of the Hat function \eqref{eq:hat_function} taken over the slice bounds, same as at first order, while the incoming coordinate is the midpoint $\bar{q}_{m,\tau} = (q_s + q_e)/2$ of the two intercepts at which the vertical line $p = \bar{p}_{m,\tau} $ cuts the triangle's edges (Fig. \ref{fig:pq_phase_space}).

\subsection{Discrete Assembly}
\label{subsec:discrete_assembly_extrapolation}

\begin{figure}[htbp]
    \centering
    \includegraphics[width=\columnwidth]{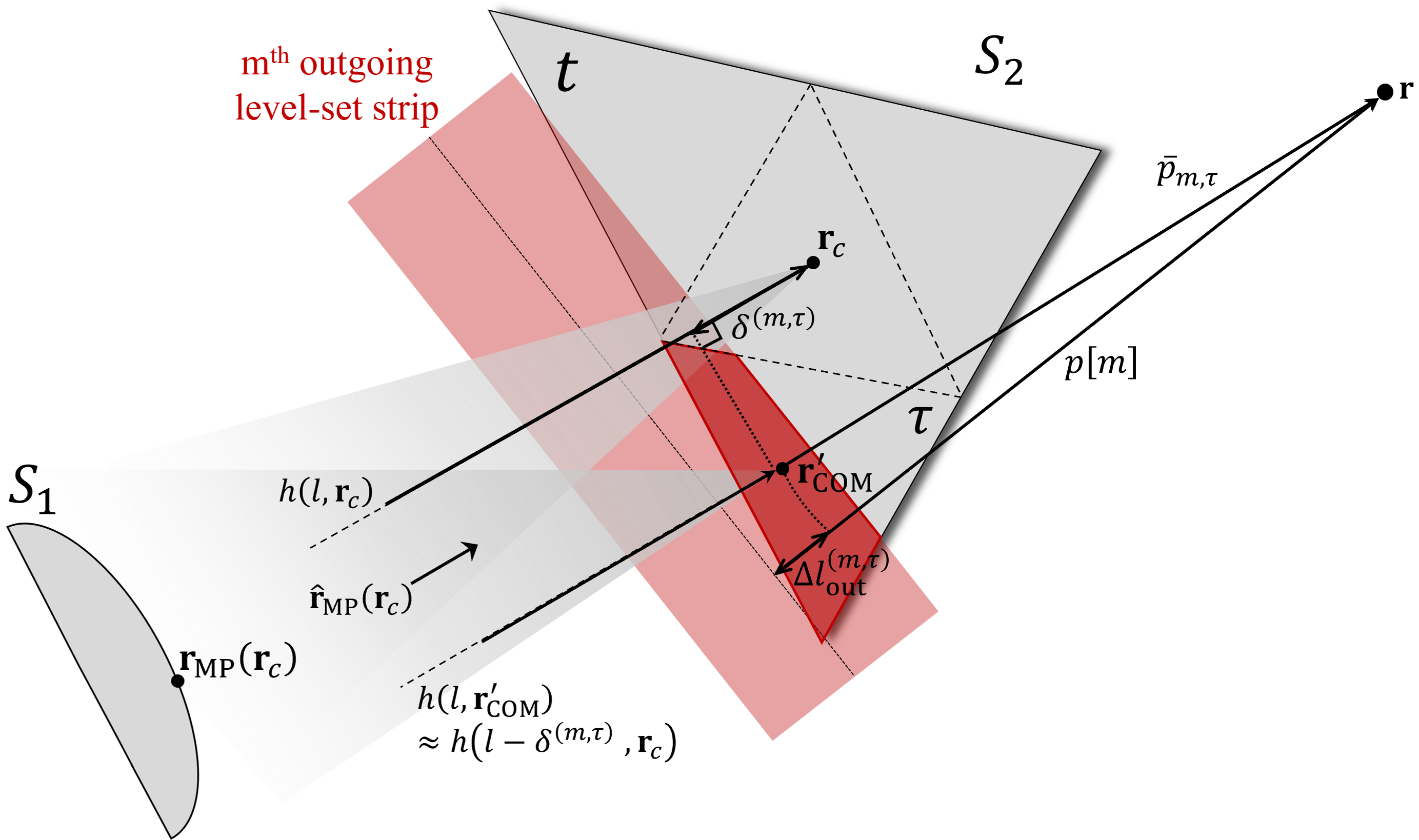}
    \caption{A field memory $h$ evaluated at a sub-triangle $\tau$ slice's centroid: the field arrives from $S_1$ along the dominant direction and the profile at $\mathbf{r}_{\text{COM}}'$ is evaluated as a translation of the profile at the centroid, then propagates by distance $\bar{p}_{m,\tau}$ toward the observer $\mathbf{r}$.}
    \label{fig:united_residuals_concept}
\end{figure}

Now that the triangles are sufficiently small, we assemble the profile over the refined set $\tau$, reading $\eta$, $K$ and the residuals from them. The three residuals of \eqref{eq:h2_resolved_assembly} are then the parent residual $\Delta l_{\text{parent}}^{(\eta,t_\tau)}$ ($t_{\tau}$ denotes the parent triangle of $\tau$); the outgoing residual $\Delta l_{\text{out}}^{(m,\tau)} = \bar{p}_{m,\tau} - p[m]$, the offset of the slice COM within its own outgoing bin; and the incoming residual $\Delta l_{\text{in}}^{(m,\tau)}$, obtained by projecting the memory from the parent's center to the slice centroid along $\hat{\mathbf{r}}_{\text{MP}}$ (Fig. \ref{fig:united_residuals_concept}). They sum to

\begin{equation}
\Delta l_{\text{res}}^{(m,\tau)} = \Delta l_{\text{parent}}^{(\eta, t_\tau)} + \Delta l_{\text{in}}^{(m,\tau)} + \Delta l_{\text{out}}^{(m,\tau)}
\label{eq:h2_total_residual_def}
\end{equation}
with the matching integer bin shift $K^{(m,\tau)}$. Evaluating \eqref{eq:h2_temporal_binning} over the refined mesh, with the resolved $I^{(m,\tau)}$ of \eqref{eq:h2_resolved_assembly} in place of $\hat{I}^{(m,t)}$, assembles the discrete second-order baseband GRP as an area-weighted, phase-corrected sum over all outgoing bins and their sub-triangles:
\begin{equation}
\begin{split}
h_B^{(ij)}[n] ={}& \frac{1}{dl} \sum_{m} \sum_{\tau \in T_{\text{AMR},m}} \left| h_B^{(i)}\big[\eta, 0\big] \right|_{\mathbf{r}_{c, t_\tau}} \\
&\times\, g^{(j)}[t_\tau]\, M_{\text{0}}[m,\tau]\, \frac{A_{\text{2D}}[\tau]}{A_{\text{2D}}[t_\tau]} \\
&\times\, e^{-j k \Delta l_{\text{res}}^{(m,\tau)}}
\end{split}
\label{eq:second_order_baseband}
\end{equation}
where $\eta = n - m - K^{(m,\tau)}$, $g^{(j)}[t_\tau]$ is the parent facet's amplitude coefficient and $A_{\text{2D}}[\tau]/A_{\text{2D}}[t_\tau]$ is the area ratio of the sub-triangle against its parent in the phase space.

\section{Wideband Frequency Extrapolation}
\label{sec:extrapolation}

Both orders assemble complex baseband profiles whose magnitudes are frequency-independent and whose phases encode a sub-bin residual extracted at the reference wavenumber $k$. Extrapolation to a target wavenumber $k'$ is therefore identical in form for the two orders, and is developed here.

The bin residual is recovered from the phase of the assembled bin, $\Delta \bar{l}_n = -\frac{1}{k}\angle\big(h_B^{(\chi)}[n]\big)$, and rescaled linearly to $\psi_n^{(k')} = -k' \Delta \bar{l}_n$. The extrapolated GRPs are
\begin{equation}
h^{(\chi)}[n, k'] = \left| h_B^{(\chi)}[n, k] \right| e^{j \psi_n^{(k')}} e^{-j k' \ell_n^{(\chi)}}
\label{eq:extrapolated_profiles}
\end{equation}
with the Baseband GRPs at wavenumber $k'$ defined as $h_B^{(\chi)}[n, k']=| h_B^{(\chi)}[n, k] | e^{j \psi_n^{(k')}}$, where $\chi$ runs over $\{(1),(2)\}$ at first order and $\{(11),(21),(12),(22)\}$ at second, with carrier path-lengths $L_n^{(\chi)} = l[n]$ in the first order and $L_n^{(\chi)} = (n+1)dl$ in the second.

Since evaluating these profiles via discrete summation treats the bins as point sources, we must account for the phase cancellation that occurs as the field is distributed across a bin. The area integration in \eqref{eq:discrete_baseband} absorbed the cancellation occurring at the reference frequency $k$; extrapolating to $k'$ requires restoring it for the new frequency via a Sinc-correction factor. For simplicity, we assume that the field is uniformly distributed across interior bins, which does not hold for an arbitrarily shaped geometry, while the leading and trailing non-empty bins are truncated by the profile's extent $[l_{\text{min}}, l_{\text{max}}]$. We therefore track an effective per-bin width,
\begin{equation}
w_n = \max\!\left(0,\ \min\!\left(l[n]+\tfrac{dl}{2}, l_{\text{max}}\right) - \max\!\left(l[n]-\tfrac{dl}{2}, l_{\text{min}}\right)\right),
\label{eq:effective_bin_width}
\end{equation}
such that $w_n = dl$ for an interior bin and $w_n \leq dl$ at the edges of the profile's range, with $[l_{\text{min}}, l_{\text{max}}]$ taken as the minimal and maximal path lengths of the field interaction with the geometry. The corresponding shape-factor correction is
\begin{equation}
S_{\text{sinc}}(k, k', w_n) = \frac{\text{sinc}(k' w_n / 2)}{\text{sinc}(k w_n / 2)}.
\label{eq:sinc_factor}
\end{equation}
This correction is meaningful only while its numerator remains inside the main lobe. The first null falls at $k' w_n = 2\pi$, and since $w_n \le dl$ the binding condition is
\begin{equation}
dl < \frac{2\pi}{k'_{\max}} = \lambda_{\min}
\label{eq:sinc_validity}
\end{equation}
where $\lambda_{\min}$ is the wavelength at the highest evaluated frequency. Equation \eqref{eq:sinc_validity} constrains the range-bin width, not the surface mesh, and it sets how far a field can be extrapolated in frequency.

Applying the correction, the wideband first-order scattered field evaluates as
\begin{equation}
\varphi^{(1)}(k') = dl \sum_{n} S_{\text{sinc}}(k, k', w_n) \left( h^{(1)}[n, k'] + j k' h^{(2)}[n, k'] \right)
\label{eq:final_discrete_field}
\end{equation}
and the second-order field as
\begin{equation}
\begin{split}
\varphi^{(2)}(k') = (dl)^2 \sum_n S_{\text{sinc}}(k, k', w_n)& \\ \times \Big[ h^{(11)}[n, k'] + j k' \Sigma_{k'}[n]
&-\, (k')^2 h^{(22)}[n, k'] \Big]
\end{split}
\label{eq:total_extrapolated_field}
\end{equation}
where $\Sigma_{k'}[n] = h^{(21)}[n, k'] + h^{(12)}[n, k']$.

Both rely on the assembled bin phase being unambiguous, which does not follow automatically from the slice residuals being bounded, since the assembled bin sums complex phasors. It does follow here because the geometric weights and scattering kernel factors are positive real numbers: a positively weighted sum of complex numbers lying in a convex cone remains in that cone \cite{boyd2004convex}, so the assembled bin phase cannot exceed the largest phase among its constituent phasors.

At first order the bin width constrains each slice residual to $\le dl/2$, bounding the constituent phases by $k\,dl/2$. At second order the residual is the sum of the three contributions of \eqref{eq:h2_total_residual_def}, each bounded by $dl/2$: the parent residual by the first-order bin limits, the incoming residual because the AMR restricts the incoming variation across a sub-triangle to $dl$, and the outgoing residual because the level-set strip is itself $dl$ wide. Hence $|\Delta l_{\text{res}}^{(m,\tau)}| \le 3dl/2$ and the aggregated phase is bounded by $3k\,dl/2$. In either case the aggregated phase remains within $\pi/2$, and hence free of wrap-around ambiguity, once $k$ and $dl$ are sufficiently small to satisfy the corresponding bound.

\section{Discrete-Time Signal Reconstruction}
\label{sec:time_domain}

Sections \ref{sec:first_order} and \ref{sec:second_order} established that the discrete baseband arrays $h_B^{(\chi)}[n]$ (with $\chi \in \{(1),(2),(11),(21),(12),(22)\}$ indexing the single- and double-bounce components) are complex-valued: each entry carries a physical geometric magnitude together with a sub-bin range residual embedded as phase. This section shows that these arrays also map onto the time-domain transient response, either through an analytic delay superposition or through a complex baseband convolution. Treating $h_B^{(\chi)}[n]$ as a standard real-valued FIR filter and convolving it against a sampled input would instead force every profile entry onto a bin center, destroying the sub-bin precision developed above.

\subsection{Analytic Time-Domain Superposition}
\label{subsec:analytic_time_superposition}

The total continuous impulse response is the superposition of the two orders already derived, $\tilde{h}(t) = \tilde{h}^{\text{I}}(t) + \tilde{h}^{\text{II}}(t)$, given by \eqref{eq:temporal_h1} and \eqref{eq:temporal_h2}. For a known input $x(t)$, the output $y(t) = x(t) * \tilde{h}(t)$ follows by convolution. Rather than differentiating $\tilde{h}(t)$ itself, we use the standard identity $x * \frac{d^m h}{dt^m} = \frac{d^m x}{dt^m} * h$ to move the time derivatives onto the input signal, and group by derivative order:
\begin{equation}
\begin{split}
y(t) = c\big[x * (h_B^{(1)} + h_B^{(11)})\big](ct) &+ \big[x' * \Sigma_B^{(1)}\big](ct) \\
&+ \tfrac{1}{c}\big[x'' * h_B^{(22)}\big](ct)
\end{split}
\label{eq:superposition_groups}
\end{equation}
where $\Sigma_B^{(1)} = h_B^{(21)} + h_B^{(12)}$, and $x'$, $x''$ are the first and second derivatives of the input.

Each array entry corresponds to a real magnitude $|h_B^{(\chi)}[n]|$ located at an exact sub-bin coordinate rather than at the bin center: $l[n]$ for single-bounce components, or $(n+1)dl$ for double-bounce components, plus the residual $\Delta\bar{l}_n^{(\chi)}$. This gives an exact fractional time delay per array entry, $\tau_n^{(\chi)} = \big(L_n^{(\chi)} + \Delta\bar{l}_n^{(\chi)}\big)/c$,
and the continuous convolutions in \eqref{eq:superposition_groups} collapse into discrete sums of \textit{delayed evaluations of the analytic input signal}. For example, the zero-derivative group becomes $y_0(t) = c\sum_n \big[dl\,|h_B^{(1)}[n]|\,x(t-\tau_n^{(1)}) + (dl)^2 |h_B^{(11)}[n]|\,x(t-\tau_n^{(11)})\big]$, with $y_1(t)$ and $y_2(t)$ built the same way from $x'$ and $x''$. This reconstruction carries the sub-bin timing precision regardless of how coarse $dl$ is; it is the time-domain counterpart of the frequency-domain extrapolation already used throughout the paper.

\subsection{The Complex Baseband Equivalent System}
\label{subsec:complex_baseband_system}

Evaluating \eqref{eq:superposition_groups} against the true passband signal is rarely practical, so radar and sonar waveforms are simulated at complex baseband. Let $x(t) = x_B(t)e^{j\omega_c t}$ for a baseband envelope $x_B(t)$ and carrier $\omega_c$. Differentiating twice and collecting terms defines the baseband derivatives $\tilde{x}_B'(t) = x_B'(t) + j\omega_c x_B(t)$ and $\tilde{x}_B''(t) = x_B''(t) + 2j\omega_c x_B'(t) - \omega_c^2 x_B(t)$, so that $x'(t) = \tilde{x}_B'(t)e^{j\omega_c t}$ and $x''(t) = \tilde{x}_B''(t)e^{j\omega_c t}$.

Applying the delay $\tau_n^{(\chi)}$ directly to this passband carrier and substituting $k_c = \omega_c/c$ gives $e^{j\omega_c(t-\tau_n^{(\chi)})} = e^{j\omega_c t}\, e^{-jk_c L_n^{(\chi)}}\, e^{-jk_c\Delta\bar{l}_n^{(\chi)}}$. The rightmost factor is exactly the bin's complex phasor at wavenumber $k_c$: multiplying it by the real magnitude $|h_B^{(\chi)}[n]|$ reconstructs the complex baseband profile $h_B^{(\chi)}[n,k_c]$ already defined in Section \ref{sec:extrapolation}, and the remaining factor of the carrier re-modulates it into the profile $h^{(\chi)}[n,k_c]$. The complex array computed for frequency-domain extrapolation therefore also serves as the convolution kernel for baseband time-domain simulation at that carrier. Substituting these into \eqref{eq:superposition_groups} and factoring out the global carrier $e^{j\omega_c t}$ yields the complex baseband output $y(t) = y_B(t)e^{j\omega_c t}$ in terms of the extrapolated GRP arrays:
\begin{align}
y_{B,0}(t) &= c\sum_n \big[dl\, h^{(1)}[n,k_c]\, x_B(t-\tau_n^{(1)}) \nonumber \\
&\quad + (dl)^2 h^{(11)}[n,k_c]\, x_B(t-\tau_n^{(11)})\big] \label{eq:baseband_y0} \\
y_{B,1}(t) &= \sum_n \big[dl\, h^{(2)}[n,k_c]\, \tilde{x}_B'(t-\tau_n^{(2)}) \nonumber \\
&\quad + (dl)^2 \Sigma_{k_c}[n]\, \tilde{x}_B'(t-\tau_n^{(12)})\big] \label{eq:baseband_y1} \\
y_{B,2}(t) &= \tfrac{1}{c}\sum_n (dl)^2 h^{(22)}[n,k_c]\, \tilde{x}_B''(t-\tau_n^{(22)}) \label{eq:baseband_y2}
\end{align}
with $y_B(t) = y_{B,0}(t) + y_{B,1}(t) + y_{B,2}(t)$ and $\Sigma_{k_c}[n]$ the same as in \eqref{eq:total_extrapolated_field}.

\section{Computational Complexity}
\label{sec:complexity}

For the first-order profile, extraction requires $\mathcal{O}(N_t N_s)$ operations and $\mathcal{O}(N_S)$ storage, where $N_t$ is the number of mesh lit triangles, $N_s$ the maximal slice count in a single triangle, and $N_S$ the total number of level-set strips sampling the geometry. Note that $N_s$ depends strongly on the phase gradient, whereas $N_S$ depends only on the target's spatial extent, $l_{max} - l_{min}$.

The second-order extraction proceeds in two stages. Assembling the field memory requires a primary profile evaluated at every secondary facet center, costing $\mathcal{O}(N_t^2 N_s)$ operations and $\mathcal{O}(N_t N_{S_1})$ memory, where $N_{S_1}$ is the maximal length of the primary profiles. This quadratic dependence on facet pairs is intrinsic to double-bounce PO and is shared with IPO; what distinguishes the present scheme is that the cost is incurred only once, at the reference frequency.

The assembly stage then costs $\mathcal{O}(N_{S_1} N_t N_A)$ operations, where $N_A$ is the maximal number of sub-triangle slices produced from a single parent facet. For a recursion capped at $L_{max}$ refinement levels a parent yields at most $4^{L_{max}}$ sub-triangles, bounding $N_A \le 4^{L_{max}} N_{s_2}$, where $N_{s_2}$ is the maximal number of outgoing-delay bins that a parent triangle intersects. The bound is loose: subdivision reduces each sub-triangle's delay span and therefore the number of bins it intersects, so refinement does not multiply the slice count by the full sub-triangle count.

The wideband evaluation is unchanged by order. Extrapolating an extracted profile to $N_f$ frequency samples requires $\mathcal{O}(N_S N_f)$ operations.

\section{Results and Numerical Validation}
\label{sec:results}

This section compares CAPS against theoretical benchmarks and against standard computational methods.

In all examples, we assume that the speed of sound in water is $1481 \text{ m/s}$, the medium around the target is homogeneous, and the target is hard (Neumann boundary conditions). The Scattering Cross Section (SCS) reported throughout this section follows the same form as the monostatic Radar Cross Section, applied here to the acoustic pressure field, $\text{SCS} = \lim_{r\to\infty} 4\pi r^2 |\varphi_s(\mathbf{r})|^2 / |\varphi_{\text{inc}}|^2$, where $\varphi_s$ is the scattered field at the observation point and $\varphi_{\text{inc}}$ is the incident field amplitude at the origin.

Two error metrics are used to compare a computed quantity $\Phi$ against a reference $\Phi_{\text{ref}}$, evaluated at discrete time or frequency samples $i$: the Normalized Root-Mean-Square Error,
\begin{equation}
\text{NRMSE} = \sqrt{\left[\tfrac{1}{N}\textstyle\sum_{i=1}^N \left(\Phi[i]-\Phi_{\text{ref}}[i]\right)^2\right] \Big/ \left[\tfrac{1}{N}\textstyle\sum_{i=1}^N \Phi_{\text{ref}}[i]^2\right]}
\label{eq:nrmse_def}
\end{equation}
and the Maximal Relative Error,
\begin{equation}
\text{MRE} = \max_{i} \left\vert{} \frac{\Phi[i] - \Phi_{ref}[i]}{\Phi_{ref}[i]} \right\vert{}.
\label{eq:relative_erorr_def}
\end{equation}
Both metrics are evaluated on the linear scale.

\subsection{Spherical Scatterer: Accuracy Against Radon Transform Physical Optics}
\label{subsec:results_sphere_radon}

To validate the computational stability and accuracy of the CAPS algorithm, it was compared against both the exact closed-form Frequency Domain Radon Transform (FD-RT) presented in \cite{Bolukbas2004Radon} and a Time Domain Radon-Transform (TD-RT) presented in \cite{Aktepe2021}. We simulated a half-unit sphere illuminated by a plane wave with $\mathbf{\hat k}_{\text{inc}} = -\mathbf{\hat z}$ in a monostatic setup. The geometry was evaluated across multiple discretization levels ranging from 1,016 to 172,968 elements on the lit side. The SCS was evaluated from $f = 49 \text{ Hz}$ to $9,088 \text{ Hz}$ with a frequency step of $\Delta f = 49 \text{ Hz}$. When calculating errors, all frequency samples were taken into account. For each discretization level, we varied the range sampling interval from $\Delta l = 0.075 \text{ m}$ down to $\Delta l = 0.18 \text{ mm}$, using the relationship $\Delta t = \Delta l / c$. The errors were calculated by comparing against the closed-form PO solution for a sphere's cross section \cite{Gordon1975, Bolukbas2004Radon}. Because the per-facet slicing of Section \ref{subsec:discrete_implementation} coincides with the Radon-transform triangle function under the far-field plane-wave illumination, this comparison isolates a single variable: integration over a bin with a phase residual, against evaluation at discrete samples. Note that this reference assumes a perfectly smooth spherical surface \cite{jenn1995radar}, whereas both solvers operate on a flat triangular model, so an additional geometric modeling error is expected.

\begin{figure}[htbp]
    \centering
    \includegraphics[width=\columnwidth]{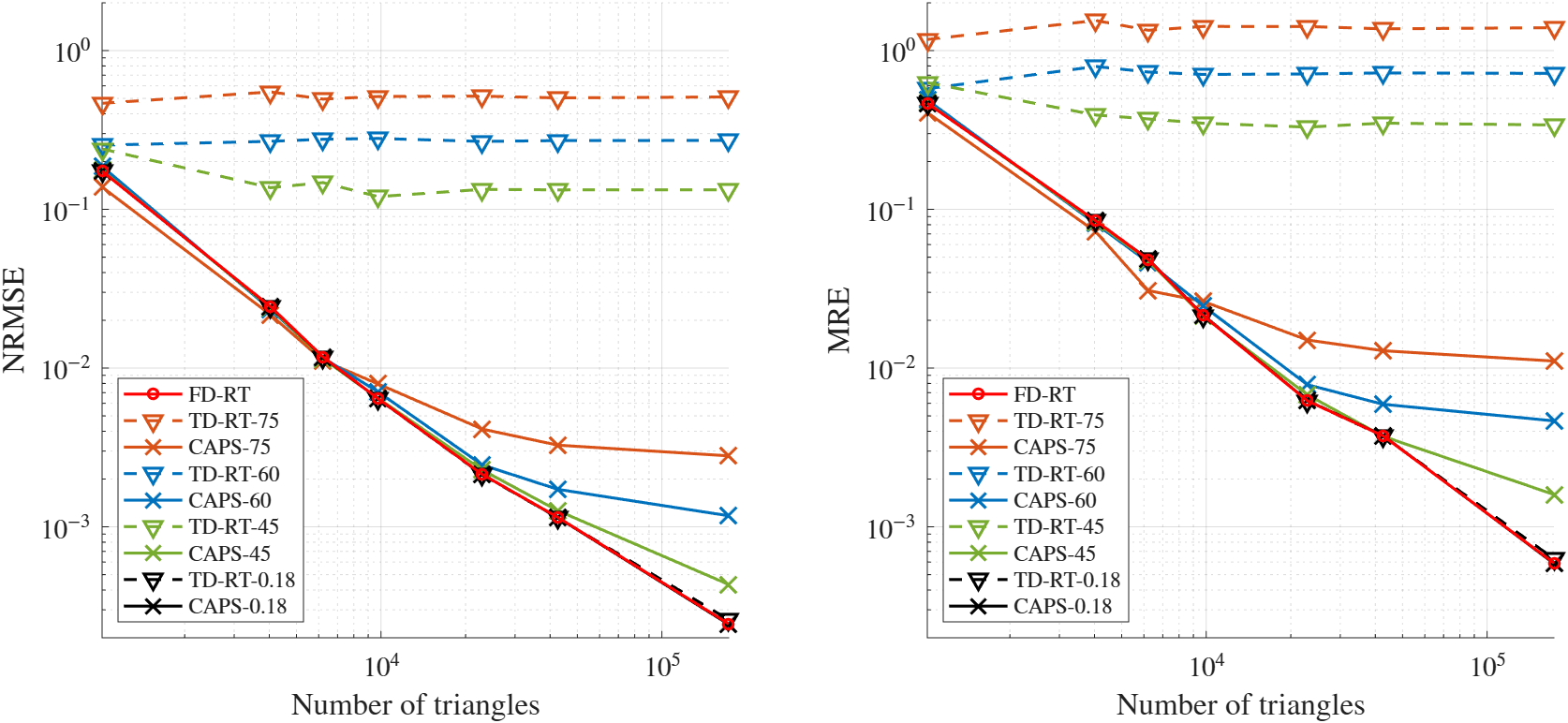}
    \caption{Scattering Cross Section NRMSE and MRE of the unit sphere for multiple range sampling resolutions.}
    \label{fig:caps_vs_rt_error}
\end{figure}

As demonstrated in Fig.~\ref{fig:caps_vs_rt_error}, standard point-sampled algorithms like TD-RT suffer from geometric omission at coarse time-sampling intervals. When the time step exceeds the physical dimensions of the mesh elements, the sampling plane fails to intersect part of the geometry, causing the error to saturate. For instance, at $\Delta l = 0.075 \text{ m}$, the TD-RT geometric omission leads to an error plateau near $50\%$, irrespective of the mesh density, indicating that the error is not caused by the mesh discretization level. TD-RT converges to the exact FD-RT solution only when a very dense time sampling is used. The number following each algorithm's label is $\Delta l$ in mm.

In contrast, the phase-residual correction lets CAPS maintain high accuracy without a dense grid. On the same coarse grid of $\Delta l = 0.075 \text{ m}$, CAPS avoids the error plateau seen in TD-RT and converges rapidly to the exact FD-RT solution.

To directly compare the convergence points of the algorithms, we calculated the sampling resolution where each algorithm achieves approximately the same NRMSE error as the FD-RT solution evaluated with 172,968 elements (i.e., $\approx 2.5 \cdot 10^{-4}$). TD-RT and CAPS converged at $\Delta l = 0.18 \text{ mm}$ and $\Delta l = 33 \text{ mm}$, respectively, a ratio of approximately 183 between the required sampling intervals.

As explained in Section \ref{sec:complexity} and in \cite{Aktepe2021}, CAPS and TD-RT share the same fundamental computational complexity. Yet CAPS converges to the exact solution with significantly coarser range sampling intervals, decreasing the required runtime and memory (i.e., decreasing $N_s$ and $N_S$, respectively). Decreasing $N_S$ also lowers the runtime needed for the wideband SCS extrapolation.

\subsection{Flat Plate in the Near-Field}
\label{subsec:results_Near_Field_flat_plate}

This example demonstrates the stability of CAPS in the near field, where it must handle spherical wavefronts. It also shows that CAPS is not strictly bounded by the mesh resolution; with a fine enough range sampling interval, it achieves stable Target Strength (TS) evaluations at frequencies that violate the standard mesh sampling rate.

For this analysis, a $2 \times 2 \times 0.1 \text{ m}$ thin flat plate is positioned at the origin, facing the $\hat{\mathbf{z}}$ direction such that its upper face rests on the $z = 0.1 \text{ m}$ plane. The geometry is discretized with flat triangles ranging from 4 to 127,416 elements on the lit side. The finest mesh yields a characteristic edge length of $1.26 \text{ cm}$. The target is illuminated by a plane wave originating from $z \rightarrow +\infty$, and two near-field observation distances are considered (relative to the plate's upper side), $Z_0 \in \{10\text{ m}, 0.5\text{ m}\}$.

A mathematical ground truth is obtained by evaluating the exact continuous geometric range profile of the plate, using the expanded impulse response formulation of Appendix \ref{app:km_generalization}. With axial plane-wave illumination the total delay reduces to $l = \sqrt{\rho^2 + Z_o^2}$, where $\rho$ is the radial distance from the plate center, so the iso-delay contours are centered circles and the integrand of the isochrone area derivative collapses to the constant $Z_o/\rho$ along each of them. Where a contour lies entirely within the plate the derivative is therefore the circumference times that integrand, $dA_L/dl = 2\pi Z_o$; where the contour is clipped by the plate edges, quadrant symmetry restricts it to an arc $[\theta_{\text{start}}, \theta_{\text{end}}]$ and gives $dA_L/dl = 4 Z_o (\theta_{\text{end}} - \theta_{\text{start}})$, with $\theta_{\text{start}} = \max(0, \arccos(x_{\text{max}}/\rho))$ and $\theta_{\text{end}} = \min(\pi/2, \arcsin(y_{\text{max}}/\rho))$. Substituting into \eqref{eq:app_h1_generalized_area} and \eqref{eq:app_h2_generalized_area} yields the exact continuous BGRPs in closed form. A standard Time-Domain Physical Optics (TDPO) solver, implemented via a piecewise constant assumption, serves as the discrete baseline for comparison.

\subsubsection{Target Strength (TS) Spectrum}
\label{subsubsec:spectral_fidelity}

\begin{figure}[htbp]
    \centering
    \includegraphics[width=\columnwidth]{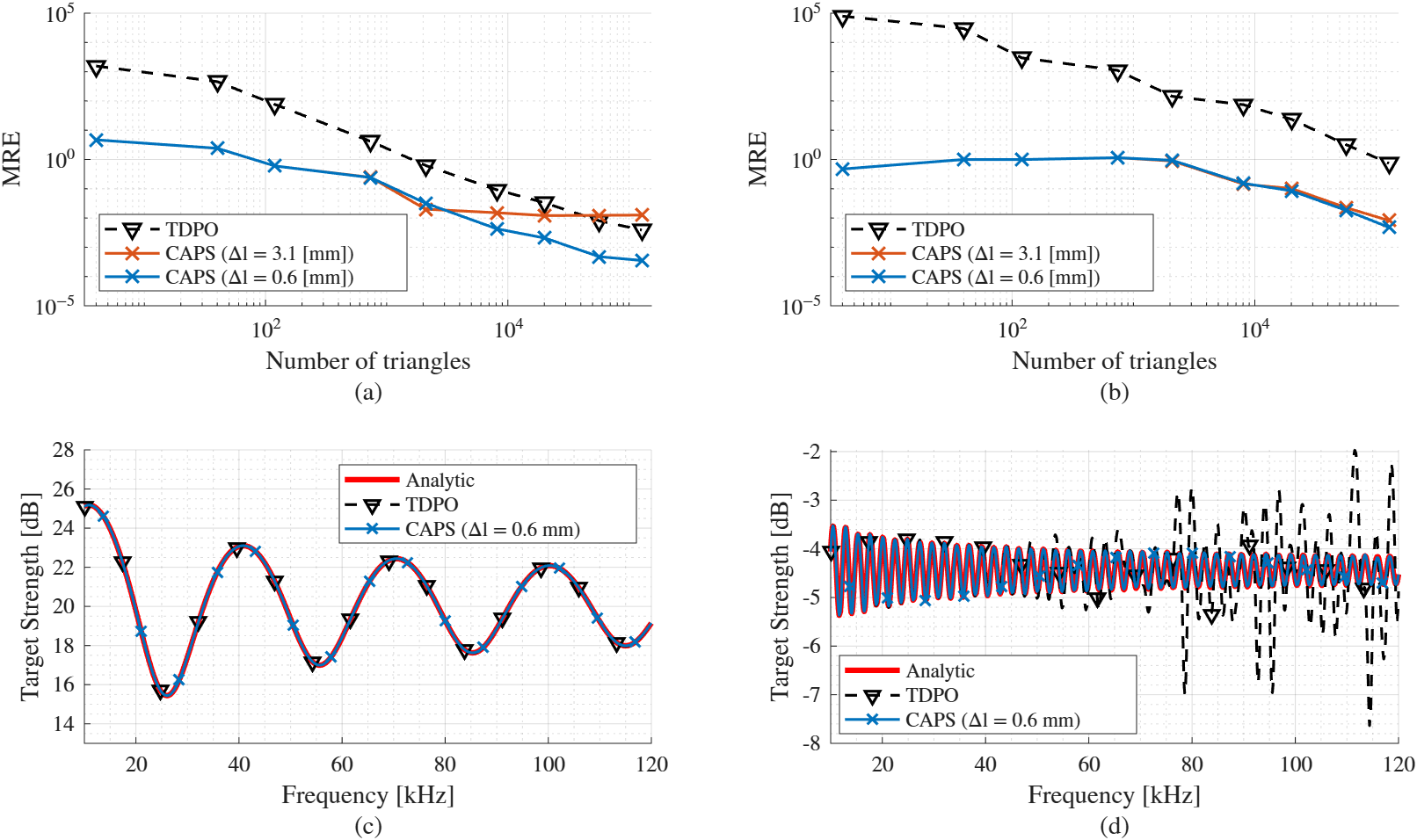}
    \caption{Near-field flat plate at two observation distances. Target Strength MRE against mesh density at (a) $Z_0 = 10\text{ m}$ and (b) $Z_0 = 0.5\text{ m}$; corresponding TS spectra against the exact analytical solution at (c) $Z_0 = 10\text{ m}$ and (d) $Z_0 = 0.5\text{ m}$.}
    \label{fig:near_field_plate}
\end{figure}

We begin by comparing the wideband acoustic TS computed by CAPS and TDPO against the exact TS from the analytical impulse response. The TS was calculated at 1,000 linearly spaced frequency samples between $10 \text{ kHz}$ and $120 \text{ kHz}$, and all frequency samples were taken into account when calculating errors. CAPS was evaluated for $\Delta l \in \{3.1, 0.6\} \text{ mm}$, where $\Delta l = 3.1 \text{ mm}$ satisfies approximately $\lambda_{\text{min}} / 4$, with $\lambda_{\text{min}}$ the wavelength at $120 \text{ kHz}$. Fig. \ref{fig:near_field_plate} reports the maximal TS error against mesh density in (a) and (b), and the corresponding TS spectrum at the finest mesh with $\Delta l = 0.6 \text{ mm}$ in (c) and (d).

At the moderate distance $Z_0 = 10 \text{ m}$, TDPO appears at first glance to outperform the coarser CAPS evaluation, reaching $0.4\%$ error against $1.2\%$ with the finest mesh. The physical target extent at this distance is only $99.5 \text{ mm}$, however, so the coarser CAPS evaluates the field from 33 non-zero bins while TDPO evaluates all 127,416 elements, a saving of roughly four orders of magnitude in memory and per-frequency runtime at a practical error level. Refining the range sampling to $\Delta l = 0.6 \text{ mm}$ reduces the CAPS error to $0.035\%$ using 165 non-zero elements, below TDPO at a fraction of the cost.

The distinction becomes clear as the observer gets closer. At $Z_0 = 0.5 \text{ m}$, TDPO reaches $74\%$ error with the finest mesh while CAPS attains $0.5\%$. The mechanism is visible in the spectra: at $10 \text{ m}$ both methods track the analytical solution closely, whereas at $0.5 \text{ m}$ the TDPO curve deviates from it at approximately $50 \text{ kHz}$, while CAPS follows the analytical solution across the entire band. The MRE curves show the same behavior. In (b) the TDPO error drops steadily with mesh size yet remains orders of magnitude above CAPS, so its limitation is the sampling interval rather than the discretization. In (a), where the two methods are closer, the coarser $\Delta l = 3.1 \text{ mm}$ evaluation plateaus at its own bin-width floor while the finer $\Delta l = 0.6 \text{ mm}$ continues to descend.

\subsubsection{Time Domain Signals}
\label{subsubsec:time_domain_signals_flat_plate_near_field}

\begin{figure}[!ht]
    \centering
    \includegraphics[width=\columnwidth]{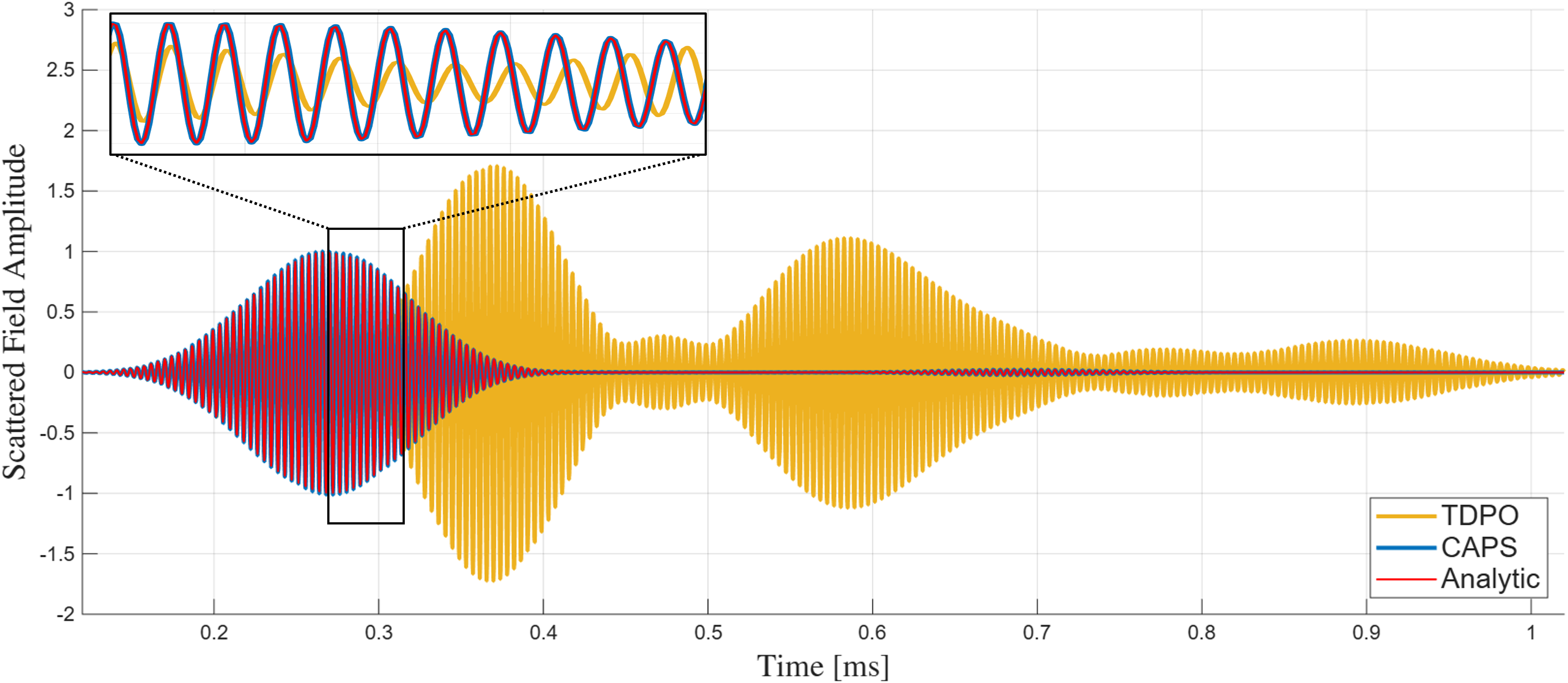}
    \caption{Scattered field amplitude with $Z_0 = 0.5\text{ m}$ for a $240\text{ kHz}$ wideband pulse.}
    \label{fig:time_signal_flat_plate_0_5_m}
\end{figure}

To visualize the performance in the time domain, the geometry is illuminated by a wideband Blackman-Harris pulse with a spatial width of $0.5 \text{ m}$ and a carrier frequency of $240 \text{ kHz}$. The plane wave originates from $z \rightarrow +\infty$, the observation distance is set to $Z_0 = 0.5 \text{ m}$ and the sampling interval to $\Delta l = 0.6 \text{ mm}$. The field's amplitude is shown in Fig.~\ref{fig:time_signal_flat_plate_0_5_m}.

TDPO produces a heavily distorted signal due to aliasing, while the CAPS waveform remains stable. CAPS therefore remains usable for transient signals at high frequencies deep in the near field, where TDPO fails on the same mesh.

\subsection{Double-Bounce Validation: Corner Reflector}
\label{subsec:results_double_bounce}

\begin{figure}[htbp]
    \centering
    \includegraphics[width=\columnwidth]{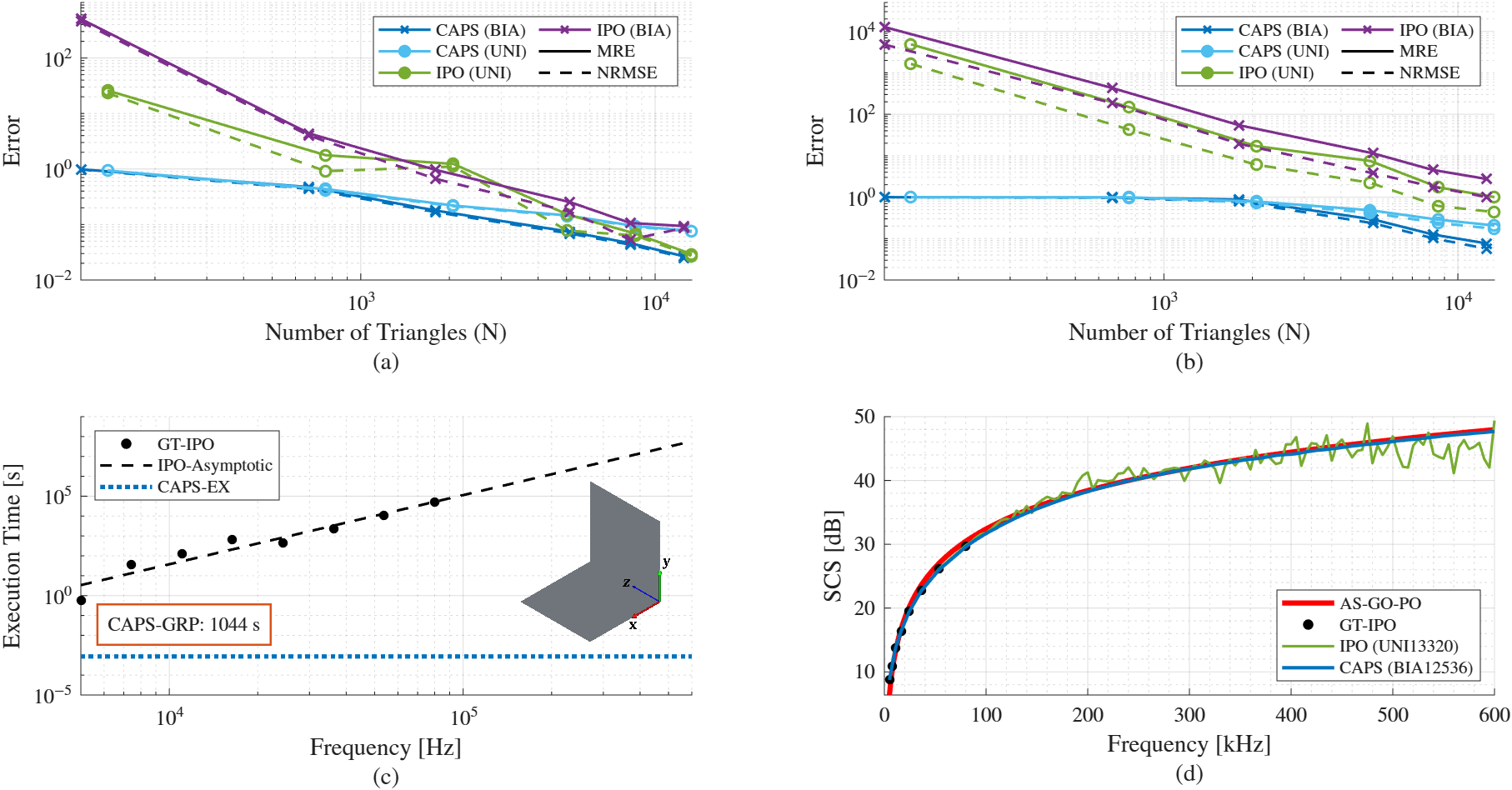}
    \caption{Double-bounce validation for a dihedral at $5^\circ$ grazing incidence. (a) Error against GT-IPO. (b) Error against AS-GO-PO. (c) Execution time vs. frequency. (d) Wideband SCS, $5$--$600\,\text{kHz}$.}
    \label{fig:dihedral_5deg}
\end{figure}

To validate the second-order formulation of Section \ref{sec:second_order}, we evaluate a two-plate dihedral under monostatic illumination at a $5^\circ$ grazing angle, isolating the double-bounce field. Fig. \ref{fig:dihedral_5deg} summarizes the validation:

\begin{itemize}
    \item \textbf{(a)} Error versus mesh size, at frequencies where a fully converged IPO run (GT-IPO) is available, for CAPS and a piecewise-constant IPO evaluation, each on a uniform (UNI) element size distribution and on a biased (BIA) mesh, reported in MRE and NRMSE (\eqref{eq:relative_erorr_def}, \eqref{eq:nrmse_def}) against GT-IPO.
    \item \textbf{(b)} The same mesh and error combinations, now compared against the asymptotic geometrical-optics solution (AS-GO-PO) \cite{Knott2004} over the $200$--$600\,\text{kHz}$ band, where GT-IPO is no longer available (motivated by (c)).
    \item \textbf{(c)} Execution time versus frequency for GT-IPO (with an asymptotic assumption of its growth), showing why (b) compares against AS-GO-PO instead at high frequency; CAPS-GRP denotes the one-time Geometric Range Profile extraction at the reference frequency, and CAPS-EX the frequency extrapolation.
    \item \textbf{(d)} The extrapolated wideband SCS from $5$ to $600\,\text{kHz}$: AS-GO-PO, GT-IPO (where available), a coarse uniform-mesh IPO run (13320 lit side elements), and CAPS on a comparably sized biased mesh (12536 lit side elements).
\end{itemize}

\begin{figure}[htbp]
    \centering
    \includegraphics[width=\columnwidth]{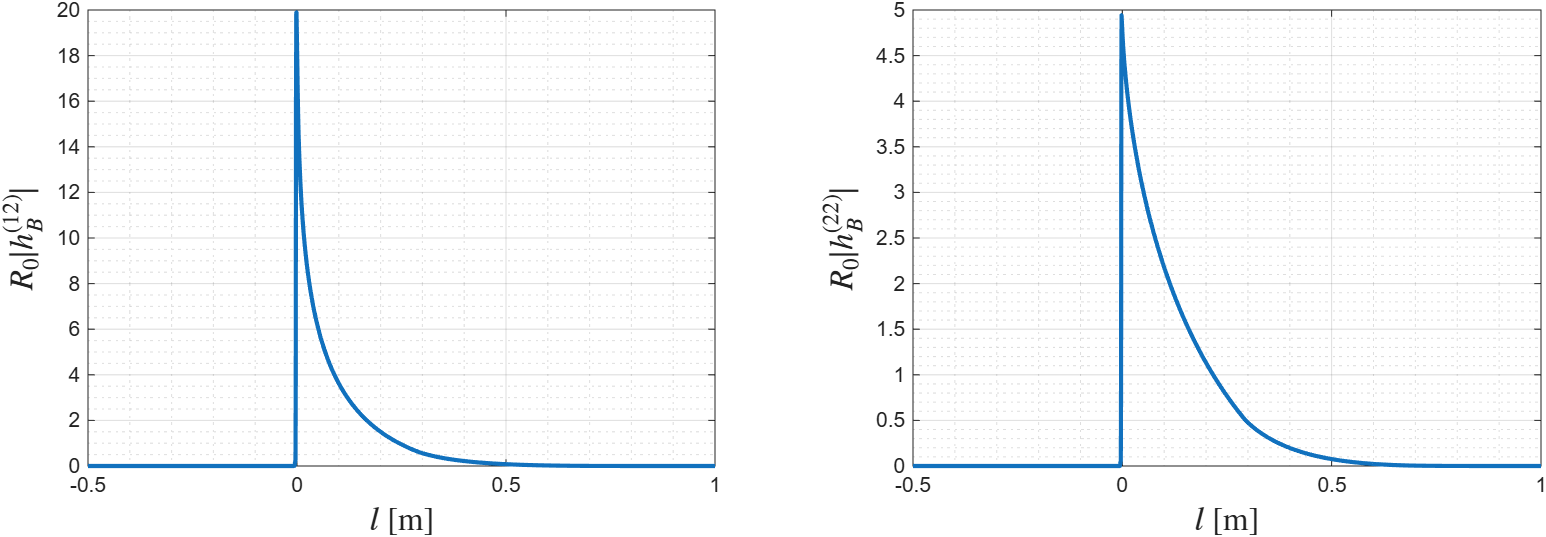}
    \caption{Second-order normalized baseband GRPs of Dihedral at $5^\circ$ grazing incidence.}
    \label{fig:dihedral_range_profiles}
\end{figure}

Two trends stand out. First, in the low-frequency band (a), CAPS on the biased mesh reaches a given error level at a substantially coarser mesh than IPO; on the uniform mesh, CAPS shows no such advantage over IPO at these frequencies. In the high-frequency band (b), by contrast, CAPS outperforms IPO on both mesh types; at the finest tested meshes, the uniform mesh for IPO and the biased mesh for CAPS (13320 and 12536 elements, respectively), the MRE is $100\%$ for IPO and $7.6\%$ for CAPS. Second, (c) and (d) together illustrate this paper's central computational claim: the one-time GRP evaluation at a reference frequency costs about the same as a single IPO run at a frequency of the same order, but every additional frequency then costs only the constant, low CAPS-EX time, whereas GT-IPO's cost grows with frequency and becomes intractable well before $600\,\text{kHz}$. In this validation case, CAPS-GRP took $1044\,\text{s}$ to compute, while GT-IPO at $80\,\text{kHz}$ alone required $50068\,\text{s}$; the average CAPS-EX time was $7.3\times10^{-4}\,\text{s}$. Extrapolating GT-IPO's own execution-time trend from (c), a single GT-IPO evaluation at $600\,\text{kHz}$ alone is expected to take approximately $58\times10^{6}\,\text{s}$. Consistent with this, (d) shows CAPS remaining smooth and closely tracking AS-GO-PO and GT-IPO across the full band, while IPO on a comparably sized mesh grows visibly noisy above roughly $180\,\text{kHz}$.

Fig.~\ref{fig:dihedral_range_profiles} presents the second-order baseband GRPs, which to our knowledge are the first of their kind. Only $h_B^{(12)}$ and $h_B^{(22)}$ are shown: the two profiles whose final bounce carries the induction kernel, $h_B^{(11)}$ and $h_B^{(21)}$, decay as $1/R^2$ and vanish in the far field. Both profiles peak at $l = 0$, since a dihedral is a retroreflector whose specular double-bounce paths share a single total length. A ray method would return that peak alone; the PO profiles additionally carry the non-specular double-bounce content, a tail extending to roughly $l = 1\,\text{m}$ from paths whose bounce points depart from the specular pair. The first bounce, between the two plates, lies in the near field, and it is this surface coupling that separates the two profiles: $h_B^{(12)}$, whose first bounce carries the induction kernel, concentrates more sharply at the junction and decays faster than $h_B^{(22)}$.

\subsection{Double-Bounce Validation: Half Ellipsoid on a Plate}
\label{subsec:results_double_bounce_ellipsoid}

To exercise stability under the gradient-\textit{magnitude} and gradient-\textit{direction} mismatch conditions discussed in Section \ref{subsec:phase_space_amr}, we run the validation on a half-ellipsoid mounted on a rectangular plate, illuminated monostatically with $\mathbf{\hat k}_\text{inc} = (-1/2 \mathbf{\ \hat x}, -1/2 \mathbf{\ \hat y}, -1/\sqrt2 \mathbf{\ \hat z})$. The half-ellipsoid has semi-axes $a = 1\,\text{m}$, $b = 0.5\,\text{m}$ and $c = 0.8\,\text{m}$ along $\hat{\mathbf{x}}$, $\hat{\mathbf{y}}$ and $\hat{\mathbf{z}}$ respectively, and rests on a $10\,\text{mm}$ thick plate spanning $x \in [-1, 2.5]\,\text{m}$ and $y \in [-0.5, 2]\,\text{m}$. The plate is sized asymmetrically so that the specular double-bounce reflection point falls inside its interior, while keeping execution time and memory down.

The high-frequency reference follows from image theory \cite{Balanis2012}: mirroring the plate completes a full ellipsoid and maps the two double-bounce paths onto bistatic returns between $\theta^i = 135^\circ$ and $\theta = 45^\circ$. Their common stationary point lies on the rim, so each path contributes half the full-ellipsoid amplitude and the two add coherently. Substituting these angles into the bistatic GO cross section of an ellipsoid \cite{Trott2007} gives
\begin{equation}
\sigma_{\text{IT-GO}} = \frac{\pi a^2 b^2 c^2}{\left(a^2\cos^2\phi + b^2\sin^2\phi\right)^2}
\label{eq:it_go_reference}
\end{equation}
which is the monostatic GO cross section of the same ellipsoid at grazing incidence \cite{Ruck1970}. This frequency-independent reference is denoted IT-GO.

\begin{figure}[htbp]
    \centering
    \includegraphics[width=\columnwidth]{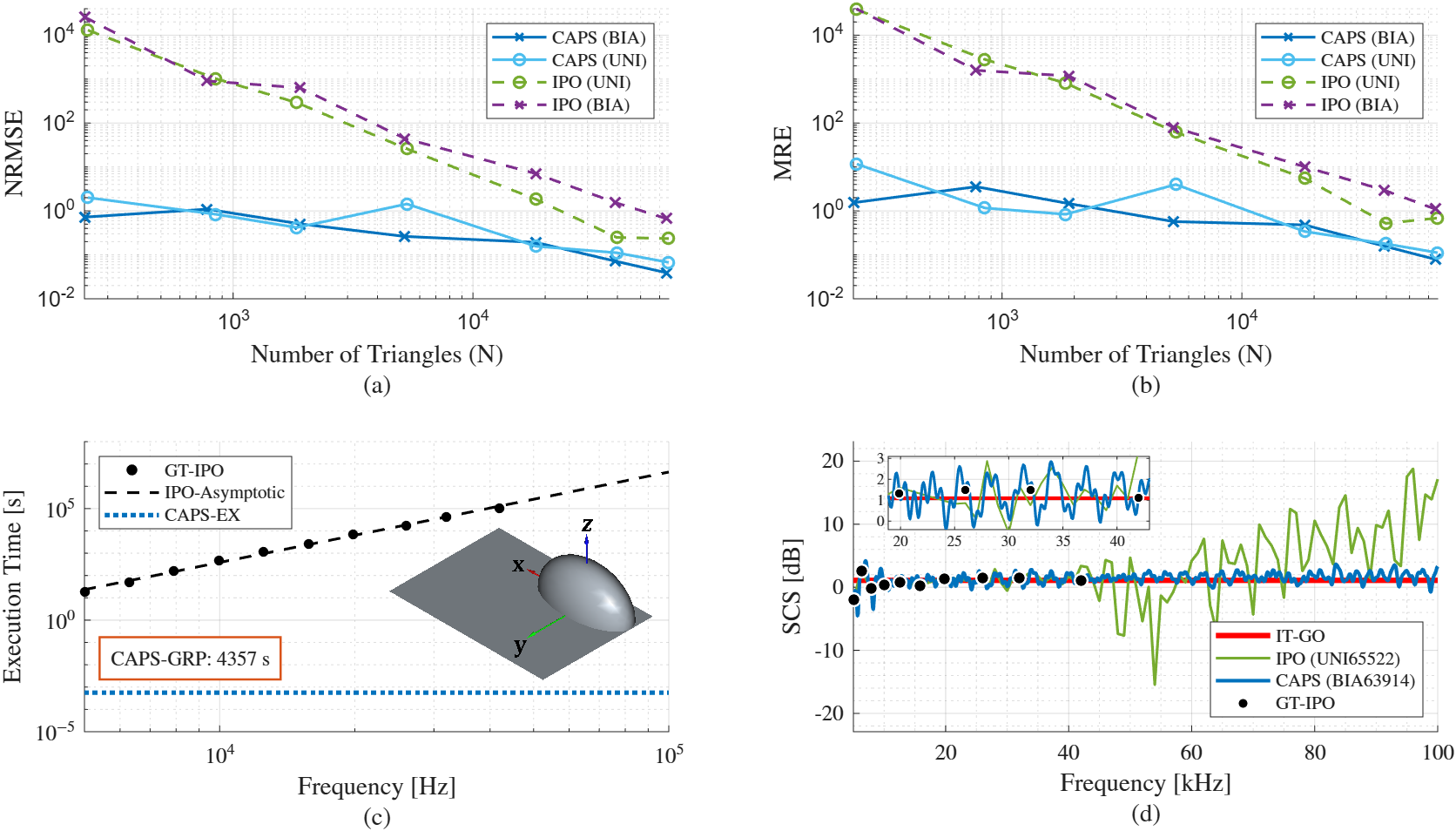}
    \caption{Double-bounce validation for a half-ellipsoid on a rectangular plate at $45^\circ$ azimuth and $45^\circ$ elevation. (a) NRMSE and (b) MRE vs.\ mesh size against GT-IPO. (c) Execution time vs.\ frequency. (d) Wideband SCS, $5$--$100\,\text{kHz}$.}
    \label{fig:ellipsoid_45_45}
\end{figure}

\begin{figure}[htbp]
    \centering
    \includegraphics[width=\columnwidth]{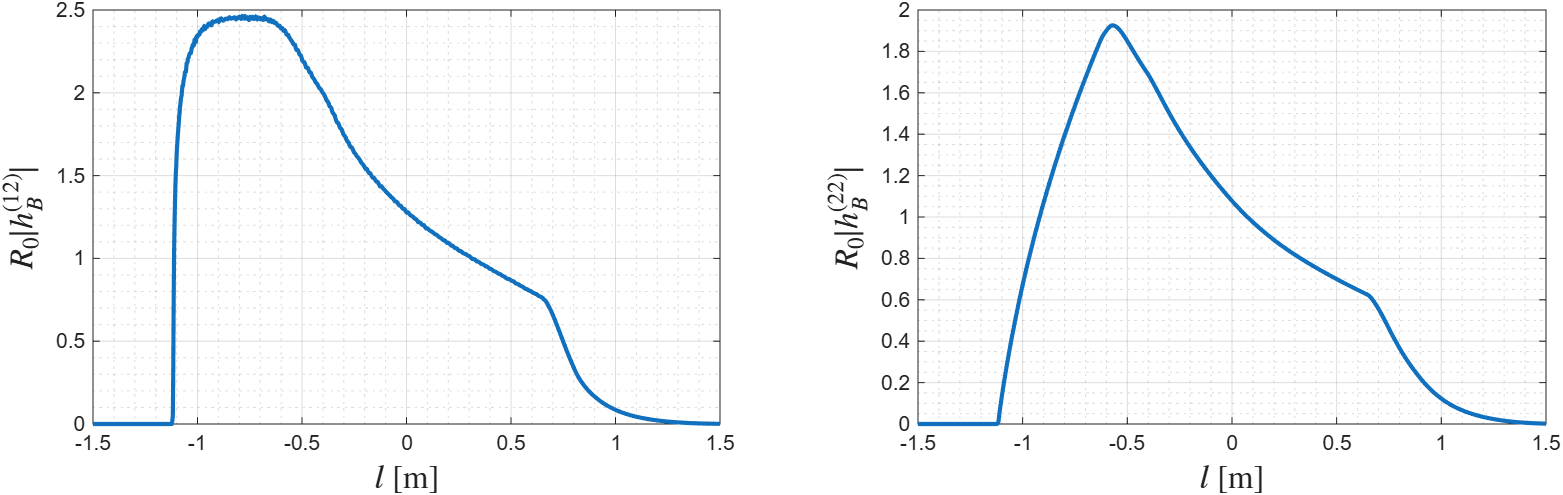}
    \caption{Second-order normalized baseband GRPs of ellipsoid on plate with $\mathbf{\hat k}_\text{inc} = (-1/2 \mathbf{\ \hat x}, -1/2 \mathbf{\ \hat y}, -1/\sqrt2 \mathbf{\ \hat z})$.}
    \label{fig:ellipsoid_range_profiles}
\end{figure}

Fig.~\ref{fig:ellipsoid_45_45} follows a four-panel layout similar to Fig.~\ref{fig:dihedral_5deg}, with NRMSE against GT-IPO in (a) and MRE against GT-IPO in (b), both only where GT-IPO is available. Two differences stand out from the Dihedral example. First, CAPS now holds a lower error compared to IPO on \textit{both} mesh types across the entire range of densities: in (a) and (b) the CAPS curve on the biased mesh remains near unity even at the coarsest mesh, whereas both IPO variants begin three to four orders of magnitude higher and converge onto CAPS only at the finest meshes tested, still yielding higher error. Second, because the primary surface is curved, the dominant arrival direction $\hat{\mathbf{r}}_{\text{MP}}$ varies from facet to facet, so the magnitude and directional gradient mismatches of Section \ref{subsec:phase_space_amr} both exist and vary across the target. The dihedral, built from two planar surfaces, presents only a fixed magnitude mismatch set by the incidence angle. The refinement depth was limited to three levels in both examples.

Panels (c) and (d) reproduce the computational picture already established for the dihedral and are therefore only summarized. GT-IPO's cost grows steeply with frequency and becomes intractable above roughly $49\,\text{kHz}$ (about $35 \text{ h}$ with a mesh of characteristic edge length $\lambda/8$), while the one-time CAPS-GRP profile extraction required $4361 \,\text{s}$ and is followed by a constant CAPS-EX cost of $6 \times 10^{-4} \,\text{s}$ per evaluated frequency. Across the $5$--$100\,\text{kHz}$ band, CAPS on a biased mesh of 63476 elements remains within a few decibels of IT-GO throughout, whereas IPO on a comparable uniform mesh of 65522 elements departs from it by more than $10\,\text{dB}$ above roughly $50\,\text{kHz}$.

Fig.~\ref{fig:ellipsoid_range_profiles} presents the second-order BGRPs of this geometry. Both begin at $l \approx -1.12\,\text{m}$, the minimum total path, corresponding to the double-bounce reflection at the rim point $(0.8944, 0.2236, 0)\,\text{m}$, which is the stationary point of the image construction above.

\section{Conclusion}
\label{sec:conclusion}

In this paper the Co-Area Phase Shifting (CAPS) algorithm is proposed, a method for extracting accurate, wideband frequency-independent Geometric Range Profiles (GRPs) from the Physical Optics surface integral via the Co-Area formula. CAPS decouples the field from the discrete mesh while tracking sub-bin phase residuals: it extracts the GRPs once at a reference frequency and then uses them for wideband frequency extrapolation. We also showed, in Section \ref{sec:time_domain}, that the complex representation of the GRPs reconstructs the transient response. For single-bounce scattering, CAPS matched the frequency-domain Radon-transform PO solution using range sampling intervals roughly two orders of magnitude coarser than a comparable time-domain Radon-transform method, and remained stable in the near field, where a standard time-domain PO solver diverges on the same coarse mesh. Single-bounce CAPS also has the same computational complexity as TD-RT.

In the double-bounce implementation, CAPS uses a surface-independent partitioning to reuse the first-order profiles as a field memory, rather than re-evaluating the single-bounce profiles at every integration point on the secondary surface. An adaptive mesh refinement then resolves the mismatch between incoming and outgoing phase gradients, refining only where needed.

The double-bounce validations show how a single range profile extraction supports wideband scattering evaluation at frequencies well above those reachable with Iterative PO.

The single-bounce CAPS has complexity of $O(N_t N_s)$ operations and $O(N_S)$ storage for evaluating the GRPs, while the double-bounce CAPS has a maximal one-time cost of $O(N_t^2 N_s)$ operations and $O(N_t N_{S_1})$ memory for the same purpose. For both orders, the per-frequency extrapolation runtime is $O(N_S)$, which does not scale with frequency.

The numerical scope presented here is restricted to piecewise-convex bodies with a single dual-interaction mechanism. Extending CAPS to general concave topologies is a next step, as is extending the discrete implementation to third and higher-order interactions using the recursive framework of Appendix \ref{app:generalization_higher_orders}.
Although the framework is presented here for an acoustic field under a Neumann boundary condition, the underlying Co-Area transformation is purely geometric, so it may be extended to vector fields and to other boundary conditions.

\appendices

\section{Phase Center-of-Mass Equivalence for Linear Gradients}
\label{app:phase_com_proof}

Section \ref{sec:second_order} approximates a slice's phase contribution by evaluating it at the slice's geometric COM rather than integrating explicitly. This appendix justifies that approximation.

Let $S$ be a small planar region of area $A$ with centroid $\mathbf{r}_c = \frac{1}{A}\iint_S \mathbf{r}\,dA$, over which a phase function $\phi(\mathbf{r}) = \mathbf{k}\cdot\mathbf{r} + \phi_0$ varies with a constant local gradient $\mathbf{k}$, the local-linearity assumption already used throughout the paper. Because $\mathbf{k}$ is constant, $\phi(\mathbf{r}_c) = \frac{1}{A}\iint_S \phi(\mathbf{r})\,dA$: the phase at the centroid equals the area-averaged phase.

Now consider the complex integral that actually appears in the discrete assembly, $Z = \iint_S e^{-j\phi(\mathbf{r})}\,dA$. Writing $\phi(\mathbf{r}) = \phi(\mathbf{r}_c) + \delta\phi(\mathbf{r})$ and factoring out the constant centroid phase gives $Z = e^{-j\phi(\mathbf{r}_c)}\iint_S e^{-j\delta\phi(\mathbf{r})}\,dA$. For a sufficiently small region, the local phase deviation satisfies $\delta\phi(\mathbf{r}) \ll \pi$ everywhere on $S$, and the first-order Taylor expansion $e^{-jx}\approx 1-jx$ applies, giving $Z \approx e^{-j\phi(\mathbf{r}_c)} \iint_S (1 - j\delta\phi(\mathbf{r}))\,dA$.
Expanding the right-hand side and substituting $\delta\phi(\mathbf{r}) = \phi(\mathbf{r}) - \phi(\mathbf{r}_c)$, the linear term vanishes exactly by the centroid identity above: $\iint_S \delta\phi\,dA = \iint_S \phi\,dA - A\,\phi(\mathbf{r}_c) = 0$. What remains is
\begin{equation}
Z \approx A\, e^{-j\phi(\mathbf{r}_c)}.
\label{eq:app_final_proof}
\end{equation}

\section{Generalization of Range Profile Extraction to Higher-Order Scattering}
\label{app:generalization_higher_orders}

To evaluate an arbitrary $p$-th order multi-bounce interaction, the geometric extraction framework is applied recursively. Since the Green's function scattering kernel at each reflection stage is decoupled into frequency-independent induction and radiation components, the number of distinct BGRPs doubles with each bounce.

To formalize this mechanism, let $\chi_p \in \{1, 2\}^p$ denote an ordered tracking sequence representing the specific combination of geometric sub-kernels applied across $p$ bounces. Because each radiation kernel ($x_m = 2$) contributes a multiplier ($jk$), the polynomial order of a given field history is set by the number of radiation kernels, $N_2(\chi_p) = \sum_{m=1}^{p} (x_m - 1)$.

Generalizing \eqref{eq:continuous_lti}, the incident field arriving at an observation point $\mathbf{r}' \in S_p$ from the preceding $p-1$ bounces is reconstructed via the 1D spatial Fourier transform of the consolidated memory profile over the accumulated intermediate delay axis $\bar{l}_{p-1}$:
\begin{equation}
\begin{split}
\varphi^{(p-1)}(\mathbf{r}') = \int &\left[ \sum_{\chi_{p-1}} (jk)^{N_2(\chi_{p-1})} h_B^{(\chi_{p-1})}(\bar{l}_{p-1}, \mathbf{r}') \right] \\
&\times e^{-j k \bar{l}_{p-1}} \, d\bar{l}_{p-1}
\end{split}
\label{eq:app_generalized_incident_field}
\end{equation}

Substituting this incident field into the generalized $p$-th order PO radiation integral and applying \eqref{eq:coarea_general_jacobian}, the 2D surface $S_p$ is mapped onto the outgoing delay axis $l_p$ using the level sets $\Gamma(l_p)$. Analogous to the double-bounce derivation, the generalized Jacobian evaluates to the surface gradient magnitude $|\nabla_{S_p} L_p(\mathbf{r}')|$, yielding:
\begin{equation}
\varphi^{(p)}(\mathbf{r}) = \int_{l_{p}} \left[ \oint_{\Gamma(l_p)} \frac{\varphi^{(p-1)}(\mathbf{r}') g_p(\mathbf{r}, \mathbf{r}', k)}{|\nabla_{S_p} L_p(\mathbf{r}')|} \, d\gamma \right] e^{-j k l_p} \, dl_p
\label{eq:app_up_coarea_applied}
\end{equation}

To map the integral onto the total delay axis we change variables to $\bar{l}_{p-1} = l - l_p$. By expanding \eqref{eq:app_generalized_incident_field} into \eqref{eq:app_up_coarea_applied}, exchanging the order of integration, and splitting the $p$-th amplitude kernel into its frequency-independent components $g_p = g_p^{(1)} + j k g_p^{(2)}$ with $g_p^{(i)}(\mathbf{r}, \mathbf{r}') = 2\,V(\mathbf{r},\mathbf{r}')\mathcal{G}^{(i)}(\mathbf{r}, \mathbf{r}')$, we isolate the purely geometric level-set integration.

This establishes the general recursive definition for generating a frequency-independent $p$-th order BGRP from the $(p-1)$-th order profile history:
\begin{equation}
h_B^{(\chi_{p-1} j)}(l) = \int_{l_p} \oint_{\Gamma(l_p)} \frac{h_B^{(\chi_{p-1})}(l - l_p, \mathbf{r}') g_p^{(j)}(\mathbf{r}, \mathbf{r}')}{|\nabla_{S_p} L_p(\mathbf{r}')|} \, d\gamma \, dl_p
\label{eq:app_recursive_h_definition}
\end{equation}
for $j \in \{1, 2\}$. 

With all $2^p$ profiles defined, the wideband $p$-th order scattered field is evaluated as a 1D Fourier transform over the total delay bounds as in \eqref{eq:continuous_lti} and \eqref{eq:final_u2_1d}. By substituting $l = c\tau$, $k = \omega/c$ and applying the inverse temporal Fourier transform ($j\omega \rightarrow \frac{d}{dt}$), the $p$-th order transient impulse response evaluates as a sum of closed-form temporal derivatives:
\begin{equation}
\tilde{h}^{(p)}(t) = \sum_{\chi_p \in \{1,2\}^p} c^{1 - N_2(\chi_p)} \frac{d^{N_2(\chi_p)}}{dt^{N_2(\chi_p)}} h_B^{(\chi_p)}(ct)
\label{eq:app_temporal_hp}
\end{equation}

This summation confirms that regardless of the number of internal bounces, the final transient field can be reconstructed by a shift-and-add cascade of one-dimensional spatial profiles.

\section{Generalization of Classical Impulse Response Theory}
\label{app:km_generalization}

Classical transient scattering theory, established by Kennaugh and Cosgriff \cite{Kennaugh1958} and adapted for acoustics by Tsao \cite{Tsao1985}, defines the backscattered far-field impulse response as proportional to the second derivative of the target's cross-sectional area. That derivation relies on flat plane-wave phase fronts, so the geometry is partitioned strictly by 2D Euclidean planes. Replacing that flat partition with the curved iso-delay contours of the Co-Area framework extends the result to the near field, and produces a second term absent from the classical law: an induction contribution proportional to the \textit{first} area derivative. The classical relation is recovered in the far-field limit, where this term decays as $1/R_o^2$ and is neglected.

\subsection{The Isochrone Projected Area}
\label{sec:app_isochrone_area}

Under point-source illumination, the Co-Area formula partitions the target geometry along curved iso-delay spatial contours (isochrones). We define an \textit{Isochrone Projected Area}, $A_L(l)$, representing the physical area enclosed by the phase front up to delay $l$, weighted by its local geometric projection toward the observer ($\hat{\mathbf{R}}$). Applying \eqref{eq:differential_area_relation}, this area is evaluated as:
\begin{equation}
A_L(l) = \int_{l_{min}}^{l} \left[ \oint_{\Gamma(l')} \frac{\hat{\mathbf{R}}(\mathbf{r}') \cdot \hat{\mathbf{n}}(\mathbf{r}')}{|\nabla_S L(\mathbf{r}')|} \, d\gamma \right] dl'
\label{eq:app_isochrone_area_def}
\end{equation}
By the Fundamental Theorem of Calculus, its spatial derivative evaluates to the inner contour integral:
\begin{equation}
\frac{dA_L(l)}{dl} = \oint_{\Gamma(l)} \frac{\hat{\mathbf{R}}(\mathbf{r}') \cdot \hat{\mathbf{n}}(\mathbf{r}')}{|\nabla_S L(\mathbf{r}')|} \, d\gamma
\label{eq:app_isochrone_derivative}
\end{equation}

\subsection{Generalized Near-Field Impulse Response}
\label{sec:app_nearfield_profiles}

Substituting the spatial kernels \eqref{eq:kernel_split} into the single-bounce profile integrals yields a contour integral modulated by the incident amplitude $|\varphi_{\text{inc}}(\mathbf{r}')|$ and the outgoing decay $|\mathbf{r} - \mathbf{r}'|$. To extract a purely geometric signature, these amplitude factors must remain constant along the contour $\Gamma(l)$.

In a monostatic setup ($\mathbf{r}_s = \mathbf{r}$), the total contour delay is $L(\mathbf{r}') = 2|\mathbf{r} - \mathbf{r}'| = l$. This forces the one-way distance to be constant ($|\mathbf{r} - \mathbf{r}'| = l/2$) along the integration path, so both amplitude terms can be factored out. Applying the identity \eqref{eq:app_isochrone_derivative}, the near-field BGRPs evaluate to spatial derivatives of the isochrone area:
\begin{align}
h_B^{(1)}(l) &= |\varphi_{\text{inc}}| \left[ \frac{2}{4\pi |\mathbf{r} - \mathbf{r}'|^2} \right] \frac{dA_L(l)}{dl} \label{eq:app_h1_generalized_area} \\
h_B^{(2)}(l) &= |\varphi_{\text{inc}}| \left[ \frac{2}{4\pi |\mathbf{r} - \mathbf{r}'|} \right] \frac{dA_L(l)}{dl} \label{eq:app_h2_generalized_area}
\end{align}

As established in \eqref{eq:temporal_h1}, the total transient field is evaluated by superimposing these spatial profiles into the temporal domain, where $l = ct$. Applying the chain rule ($\frac{d}{dt} = c \frac{d}{dl}$), the spatial area derivatives convert into time derivatives, yielding the generalized near-field transient response:
\begin{equation}
\begin{split}
\tilde{h}^{\text{I}}(t) ={}& |\varphi_{\text{inc}}| \left[ \frac{2}{4\pi |\mathbf{r} - \mathbf{r}'|^2} \right] \frac{dA_L(ct)}{dt} \\
&+ |\varphi_{\text{inc}}| \left[ \frac{2}{4\pi c |\mathbf{r} - \mathbf{r}'|} \right] \frac{d^2A_L(ct)}{dt^2}
\end{split}
\label{eq:app_generalized_impulse_final}
\end{equation}

\subsection{Asymptotic Far-Field Evaluation}
\label{sec:app_far_field_KM}

We now recover the classical limit. An ideal plane wave ($|\varphi_{\text{inc}}| = 1$) illuminates the target along $+\hat{\mathbf{z}}$, with an observer at $(0,0,-R_o)$ and $R_o \gg z$, so $|\mathbf{r} - \mathbf{r}'| \approx R_o$. As $R_o \to \infty$ the induction term of \eqref{eq:app_generalized_impulse_final} decays as $1/R_o^2$ and is neglected, leaving
\begin{equation}
\tilde{h}^{\text{I}}(t) \Big|_{R_o \rightarrow \infty} = \frac{1}{2\pi R_o c} \frac{d^2A_L(ct)}{dt^2}
\label{eq:app_ff_impulse_response}
\end{equation}
With $A(z)$ the physical cross-sectional area at depth $z$, total path $l = 2z + R_o$ and delay $t_o = R_o/c$, the mapped Co-Area function aligns with the classical parameterization at the localized time $t' = t - t_o$: $A_L(ct) = A\left(c(t - t_o)/2\right) = S(t')$. The constant delay vanishes under the second temporal derivative, so \eqref{eq:app_ff_impulse_response} becomes $\tilde{h}^{\text{I}}(t)|_{R_o \to \infty} = (2\pi R_o c)^{-1} d^2 S(t')/dt'^2$, which is the transient scattered field of Kennaugh and Cosgriff. Factoring out the $1/R_o$ spherical divergence isolates the distance-invariant geometric amplitude and recovers Tsao's normalized acoustic impulse response for rigid boundaries \cite{Tsao1985}:
\begin{equation}
P_I(t') = \frac{1}{2\pi c} \frac{d^2 S(t')}{dt'^2} = R_o \tilde{h}^{\text{I}}(t) \Big|_{R_o \rightarrow \infty}
\label{eq:app_acoustic_alignment}
\end{equation}
The Co-Area slicing framework therefore contains classical far-field transient theory as its asymptotic limit.

\section*{Acknowledgment}
The authors would like to thank Prof. E. Heyman for insightful discussions.

The core idea, the main derivations and the novel contributions are the authors’ own. Generative AI tools were used to draft the text of the article from the first author’s M.Sc. thesis and to revise its language. The authors reviewed all text and code and take full responsibility for the content of this article.

\ifCLASSOPTIONcaptionsoff
  \newpage
\fi

\begin{IEEEbiography}{Guy Binderman}
received the B.Sc. degree in electrical engineering from Tel Aviv University, Tel Aviv, Israel, in 2018, where he is currently pursuing the M.Sc. degree with the School of Electrical Engineering.

\end{IEEEbiography}

\begin{IEEEbiography}{Amir Boag}
(S’89–M’91–SM’96–F’08) received the
B.Sc. degree (summa cum laude) in electrical engineering
and the B.A. degree (summa cum laude) in
physics, and the M.Sc. and Ph.D. degrees in electrical
engineering from the Technion—Israel Institute
of Technology, Haifa, Israel, in 1983, 1985, and
1991, respectively.

From 1991 to 1992, he was on the faculty of
the Department of Electrical Engineering, Technion.
From 1992 to 1994, he was a Visiting Assistant
Professor with the Electromagnetic Communication
Laboratory, Department of Electrical and Computer Engineering, University of
Illinois at Urbana–Champaign, Champaign, IL, USA. In 1994, he joined Israel
Aircraft Industries, Ben Gurion Airport, Israel, as a Research Engineer and
became the Manager of the Electromagnetics Department in 1997. Since 1999,
he has been with the Physical Electronics Department, School of Electrical
Engineering, Tel Aviv University, Tel Aviv, Israel, where he is currently
a Professor. He has authored or co-authored over 100 journal articles and
presented over 200 conference papers on electromagnetics and acoustics.
His current research interests include computational electromagnetics, wave
scattering, imaging, and the design of antennas and optical devices.
\end{IEEEbiography}

\end{document}